\documentclass[a4paper,12pt]{article}

\usepackage{amsmath}
\usepackage{amsfonts}
\usepackage{amssymb}
\usepackage{graphicx}
\newcommand{\de}{\partial}
\def\be{\begin{equation}}
\def\ee{\end{equation}}

\newcommand{\del}{\delta}

\newcommand{\Om}{\Omega}
\newcommand{\ud}{{\rm d}}
\newcommand{\rmd}{\textrm{d}}
\newcommand{\nab}{\nabla}

\newcommand{\m}{{m}}

\begin{document}
\def\thefootnote{\fnsymbol{footnote}}

\begin{center}
\Large{\textbf{Spherical collapse in quintessence models\\
 with zero speed of sound}} \\[0.5cm]
 
\large{Paolo Creminelli$^{\rm a}$, Guido D'Amico$^{\rm b,c}$,
Jorge Nore\~na$^{\rm b,c}$, \\[.1cm] Leonardo Senatore$^{\rm d}$ and Filippo Vernizzi$^{\rm e}$}
\\[0.5cm]

\small{
\textit{$^{\rm a}$ Abdus Salam International Centre for Theoretical Physics\\ Strada Costiera 11, 34151, Trieste, Italy}}

\vspace{.2cm}

\small{
\textit{$^{\rm b}$ SISSA, via Beirut 2-4, 34151, Trieste, Italy}}

\vspace{.2cm}

\small{
\textit{$^{\rm c}$ INFN - Sezione di Trieste, 34151 Trieste, Italy}}

\vspace{.2cm}

\small{
\textit{$^{\rm d}$ School of Sciences, Institute for Advanced Study,
  \\ Olden Lane, Princeton, NJ 08540, USA}}

\vspace{.2cm}

\small{
\textit{$^{\rm e}$ CEA, IPhT, 91191 Gif-sur-Yvette c\'edex, France\\ CNRS, URA-2306, 91191 Gif-sur-Yvette c\'edex, France}}

\end{center}

\vspace{.8cm}

\hrule \vspace{0.3cm}
\noindent \small{\textbf{Abstract}\\
We study the spherical collapse model in the presence of quintessence with negligible speed of sound. This case is particularly motivated for $w<-1$ as it is required by stability. As pressure gradients are negligible, quintessence follows dark matter during the collapse. The spherical overdensity behaves as a separate closed 
FLRW universe, so that its evolution can be studied exactly. We derive the critical overdensity for collapse and we use the extended Press-Schechter theory to study how the clustering of quintessence affects the dark matter mass function. The effect is dominated by the modification of the linear dark matter growth function. A larger effect occurs on the total mass function, which includes the quintessence overdensities. Indeed, here quintessence constitutes a third component of virialized objects, together with baryons and  dark matter, and contributes to the total halo mass by a fraction $\sim (1+w) \Omega_Q/\Omega_m$.  This gives a distinctive modification of the total mass function at low redshift.} \\
\noindent
\hrule
\def\thefootnote{\arabic{footnote}}
\setcounter{footnote}{0}

\section{Introduction}

One of the most important open questions for cosmology is whether dark energy is a dynamical component of the universe or a cosmological constant. A plethora of experiments and future observations are currently planned with the aim of improving our understanding of this question (see for instance \cite{Albrecht:2006um} for a review). One of the most popular models of dynamical dark energy is quintessence, where the acceleration of the universe is driven by a scalar field with negative pressure. Standard quintessence is described by a minimally coupled canonical scalar field \cite{Zlatev:1998tr}. In this case, scalar fluctuations propagate at the speed of light and sound waves maintain quintessence homogeneous on scales smaller than the horizon scale \cite{Ferreira:1997au}. Quintessence clustering takes place only on scales of order the Hubble radius,  so that its effect is strongly limited by cosmic variance.  However, quintessence modifies the growth evolution of dark matter through its different expansion history. Thus, future galaxy catalogs and weak lensing surveys will have a great potential in constraining the dark energy properties, in particular its equation of state, through a detailed study of the evolution of dark matter.

The study of quintessence perturbations -- where by quintessence we indicate a dark energy sector described by a single scalar degree of freedom -- has been recently revived in \cite{Creminelli:2008wc}. Here the most general theory of quintessence perturbations around a given background was derived using the tools developed in \cite{Creminelli:2006xe,Cheung:2007st}, formulated in the context of an effective field theory. An important conclusion is that quintessence with an equation of state $w < -1$ can be free from ghosts and gradient instabilities \cite{Creminelli:2006xe,Creminelli:2008wc}.  In this regime, stability can be guaranteed by the presence of higher derivative operators \cite{ArkaniHamed:2003uy,Creminelli:2006xe} with the requirement that the speed of sound of propagation is extremely small, $|c_s| \lesssim 10^{-15} $. On cosmological scales, higher derivative operators are phenomenologically irrelevant and quintessence simply behaves as a perfect fluid with negative pressure but practically zero speed of sound \cite{Creminelli:2008wc}. 
Interestingly, this description applies also in the non-linear regime,  i.e.~when perturbations in the quintessence energy density become non-linear,  as long as the effective theory remains valid.

This study motivates the possibility that quintessence has a practically zero speed of sound. Apart from these theoretical considerations, the fact that the speed of sound of quintessence may vanish opens up new observational consequences. Indeed, the absence of quintessence pressure gradients allows instabilities to develop
on all scales, also on scales where dark matter perturbations become non-linear. 
Thus, we expect quintessence to modify the 
growth history of dark matter not only through its different background evolution but also by actively participating to the structure formation mechanism, in the linear and non-linear regime, and by contributing to the total mass of virialized halos.

In the linear regime, a series of articles have investigated the observational consequences of a clustering quintessence.
In particular, they have studied the different impact of quintessence with $c_s=1$ or $c_s=0$ on the cosmic microwave background \cite{DeDeo:2003te,Weller:2003hw,BeanDore,Hannestad:2005ak}, galaxy
redshift surveys \cite{Takada:2006xs}, large neutral hydrogen surveys \cite{TorresRodriguez:2007mk}, or on the cross-correlation of the integrated Sachs-Wolfe effect in the cosmic microwave background with large scale structures \cite{Hu:2004yd,Corasaniti:2005pq}.  On non-linear scales, the dependence of the dark matter clustering on the equation of state of a homogeneous quintessence, i.e.~with $c_s=1$, has been investigated using N-body simulations in a number of articles (see for instance \cite{Alimi:2009zk} and references therein for a recent account). 

A popular analytical approach to study non-linear clustering of dark matter without recurring to N-body simulations is the spherical collapse model \cite{Gunn:1972sv}. In this approach, one studies the collapse of a spherical overdensity and determines its critical overdensity for collapse as a function of redshift. Combining this information with the extended Press-Schechter theory \cite{Press:1973iz,Bond:1990iw} one can provide a statistical model for the formation of structures which allows to predict the abundance of virialized objects as a function of their mass. Although it fails to match the details of N-body simulations, this simple model works surprisingly well and can give useful insigths into the physics of structure formation. Improved models accounting for the complexity of the collapse exist in the literature and offer a better fit to numerical simulations. For instance, it was shown in \cite{Sheth:1999mn} that a significant improvement can be obtained by considering an ellipsoidal collapse model. See also \cite{Maggiore:2009rv,Maggiore:2009rw} for recent theoretical developments and new improvements in the excursion set theory.

The spherical collapse can be generalized to include a cosmological constant (see for instance \cite{Lahav:1991wc}) and quintessence with  $c_s=1$ \cite{Wang:1998gt} (see also \cite{Weinberg:2002rd,Battye:2003bm} for subsequent applications). If quintessence propagates at the speed of light it does not cluster with dark matter but remains homogeneous. Indeed, pressure gradients contribute to maintain the same energy density of quintessence between the inner and outer part of the spherical overdensity. A study of the spherical collapse model with different quintessence potentials was performed in \cite{Mota:2004pa}. For a nice review on structure formation with homogeneous dark energy see also \cite{Percival:2005vm}.

In this paper we study the spherical collapse model in the case of quintessence with zero speed of sound. This represents the natural counterpart of the opposite case $c_s=1$.
Indeed, in both cases there are no characteristic length scales associated to the quintessence clustering\footnote{The characteristic length scale associated to the quintessence clustering is the sound horizon scale, i.e., $L_s \equiv a \int c_s dt/a$. As mentioned above, this vanishes for $c_s=0$ so that clustering takes place on all scales. For $c_s=1$ we have $L_s = 2 H_0^{-1}$, which is much larger than the scales associated to the spherical collapse.}  and the spherical collapse remains independent of the size of the object.
For this study we describe quintessence using the model developed in \cite{Creminelli:2008wc}. As explained, this description remains valid also when perturbations become non-linear and can thus be applied to the spherical collapse model. 

As the spherical collapse occurs on length scales much smaller than the Hubble radius, we will describe it using a convenient coordinate system where the effect of the Hubble expansion can be treated as a small perturbation to the flat spacetime. In these ``local'' coordinates the description of the spherical collapse becomes extremely simple and the well-known cases can be easily extended to quintessence with $c_s=0$. In this case pressure gradients are absent and quintessence follows dark matter during the collapse.
Thus, in contrast with the non-clustering case $c_s=1$ where quintessence and dark matter are not comoving, for $c_s=0$ the collapsing region is described by an exact 
Friedmann-Lema\^{i}tre-Robertson-Walker (FLRW) universe.
Note that even though the energy density of quintessence develops inhomogeneities as long as the collapse proceeds, the pressure inside and outside the overdense region remains the same. Thus, as explained below, our model does not give the same description of clustering quintessence as that proposed by \cite{Mota:2004pa} and studied, for instance, in \cite{Maor:2005hq,Manera:2005ct,Nunes:2004wn}. 

We will see that, besides quantitative differences with respect to the $c_s=1$ case -- a different threshold for collapse and a different dark matter growth function -- the $c_s=0$ case has a remarkable qualitatively new feature. Quintessence clusters together with dark matter and participates in the total mass of the virialized object, contributing to their gravitational potential.

The plan of the paper is the following. In section \ref{sec:model} we describe quintessence models with $c_s=0$. In section \ref{sec:local_coords} we study spherical collapse solutions first in known cases (dark matter only, $\Lambda$CDM and $c_s=1$ quintessence) and then in the case of a clustering dark energy,  $c_s=0$. It turns out to be much simpler to describe these solutions in coordinates for which the metric is close to Minkowski around a point in space. The equation for the evolution of the spherical collapse are solved in section \ref{sec:solving} and the threshold for collapse is calculated in the various cases. This leads to the calculation of the dark matter mass function in section \ref{sec:massfunction}. In section \ref{sec:quintmass} we study the accretion of quintessence to the dark matter haloes and its effect on the total mass function. The contribution of quintessence to the mass may be distinguished from the dark matter and baryon component in cluster measurements, as we briefly discuss in section \ref{sec:3masses}. Conclusions and future directions are discussed in section \ref{sec:conclusions}. 

\section{The model: quintessence with $c_s^2=0$}
\label{sec:model}

Let us consider a $k$-essence field described by the action \cite{ArmendarizPicon:1999rj,ArmendarizPicon:2000dh}
\begin{equation}
S = \int \ud^4 x \sqrt{-g} \; P(\phi,X) \;,
\qquad X = - g^{\mu \nu} \de_\mu \phi \de_\nu \phi \;.
 \label{kaction}
\end{equation}
The evolution equation of $\phi$ derived from this equation is
\be
 \frac1{ \sqrt{-g}} \partial_\mu (\sqrt{-g} \; 2 P_{,X} \partial^\mu \phi ) = - P_{,\phi}\;,
\label{phi_evolution}
\ee
where $P_{,f} \equiv \partial P/\partial f$.
The energy-momentum tensor of this field can be derived using 
\be
T_{\mu \nu} = - \frac{2}{\sqrt{-g}} \frac{\del S}{\del g^{\mu \nu}} \;,
\ee
and can be written in the perfect fluid form as \cite{Garriga:1999vw}
\be
T_{\mu \nu} = (\rho_Q + p_Q) u_\mu u_\nu + p_Q g_{\mu \nu}\;,
\ee
once we identify
\be 
\rho_Q=2 P_{,X} X - P \;, \qquad p_Q=P\;, \qquad u_\mu = - \frac{\partial_\mu \phi}{\sqrt{X}}\;.
\label{rho_P_u}
\ee

Let us initially neglect perturbations of the metric and assume a flat FLRW universe with metric $\ud s^2 = - \ud t^2 + a^2(t) \ud {\vec x}^2$.  
The energy-momentum tensor of the field can be perturbed around a given background solution
$\bar \phi(t)$ corresponding to a background energy density and pressure,
\be
\bar \rho_Q=2 \bar P_{,X} \bar X -\bar P \;, \qquad \bar p_Q= \bar P\;,
\label{backrho_p}
\ee
where $\bar X = \dot{\bar \phi}^{2}$.
To describe perturbations it is useful to write the scalar field as \cite{Creminelli:2006xe,Creminelli:2008wc}
\begin{equation}
 \phi(t,\vec x) = \bar \phi(t + \pi(t,\vec x)) \;,
\end{equation}
where $\pi$ describes the difference between the uniform time and scalar field hypersurfaces.\footnote{We assume that $\bar \phi$ is a monotonous function of $t$.}
Then, eq.~(\ref{rho_P_u}) can be expanded linearly in $\pi$ using 
$\phi(t, \vec x) = \bar \phi + \dot{\bar  \phi} \; \pi$ and $X(t, \vec x)  = \bar  X + \dot{\bar X} \pi + 2 \bar X \dot \pi$. This yields, for the perturbations of the energy density, pressure and velocity, 
\be
\delta \rho_Q=\dot{\bar \rho}_Q \pi +(\bar \rho_Q + \bar p_Q +4 M^4) \dot \pi \;, \qquad \delta p_Q=\dot {\bar p}_Q \pi +(\bar \rho_Q +\bar  p_Q) \dot \pi \;, \qquad u_i = - \partial_i \pi\;,
\label{deltarho_p}
\ee
where we have used eq.~(\ref{backrho_p}) 
and defined $M^4 \equiv \bar P_{,XX} \bar X^2$, where $M$ has the
dimension of a mass.

To describe the evolution of perturbations we can expand the action
(\ref{kaction}) up to second order in $\pi$ as done in \cite{Creminelli:2008wc},
\begin{equation}
  S  = \!\int \!\ud^4 x \,a^3
  \left[ \dot{\bar P} \pi + 2 \bar P_{,X} \bar X \dot{\pi}+    
    \left( \bar P_{,X} \bar X + 2 \bar P_{,XX} \bar X^2 \right) \dot{\pi}^2
  - \bar P_{,X} \bar X \frac{(\vec \nab \pi)^2}{a^2}     +  \frac{1}{2} \ddot{\bar P} \pi^2 
  + 2 \left(\bar P_{,X} \bar X \right)\dot{} \, \pi \dot{\pi} 
  \right] \; .
\label{kessence2nd}
\end{equation}
The second term proportional to $\dot \pi$ can be integrated by parts so that the part of the action linear in $\pi$ can be written using eq.~(\ref{backrho_p}) as  $- \left[ \dot {\bar \rho}_Q + 3 H (\bar \rho_Q+\bar p_Q) \right] \pi$, where $H\equiv \dot a /a$ is the Hubble rate. This part cancels due to the background equation of motion.  Furthermore, we can manipulate the last two terms of the action 
integrating by parts the last term, proportional to $\pi \dot \pi$,  and making use of
the background equation of motion, to rewrite them as $ 3 \dot{H} \bar P_{,X} \bar X \, \pi^2$.
Finally, it is convenient to rewrite the coefficients left in this expansion in terms of the background energy density and pressure using eq.~(\ref{backrho_p}). This yields
\begin{equation}
  \label{kessence2}
  S = \!\int \!\ud^4 x \, a^3
\Bigg[ \frac{1}{2} \left( \bar \rho_Q+\bar p_Q +4 M^4  \right) \dot{\pi}^2
  - \frac{1}{2}(\bar \rho_Q+\bar p_Q) \frac{(\vec \nab \pi)^2}{a^2}
  + \frac{3}{2} \dot{H} (\bar \rho_Q+\bar p_Q)  \pi^2  \Bigg] \; .
\end{equation}

The coefficients of this quadratic action are completely specified by the background 
quantities $\bar \rho_Q +\bar p_Q$ and $M^4$. The latter is a function of time which we expect to vary slowly with a rate of order Hubble.\footnote{The time variation of $M^4$ is expected to be even slower than Hubble, i.e.~of order $(1+w) H$, which is the typical time variation of $\rho_Q$.} As shown in \cite{Cheung:2007st,Creminelli:2008wc}, eq.~(\ref{kessence2}) is the most general action describing quintessence in absence of operators with higher-order spatial derivatives. Note that this action is even more general than the starting Lagrangian (\ref{kaction}) as it can be generically derived using only symmetry arguments \cite{Cheung:2007st}. 
An advantage of eq.~(\ref{kessence2}) is that its coefficients are written in terms of observable quantities. Indeed, $\bar \rho_Q +\bar p_Q$ is proportional to $1+w$, where $w \equiv \bar p_Q/\bar \rho_Q$ is the equation of state of quintessence, which we will take here and in the following to be constant.
The parameter $M^4$ is related to the speed of sound of quintessence, given by
\be
c_s^2 = \frac{\bar \rho_Q +\bar p_Q}{\bar \rho_Q +\bar p_Q +4 M^4}\;.
\label{speedsound}
\ee
As can be seen from this equation, absence of ghost -- i.e., positiveness of the time kinetic-term in eq.~(\ref{kessence2}) -- implies that $c_s^2$ has the same sign as $1+w$ \cite{Hsu:2004vr,Creminelli:2006xe,Creminelli:2008wc}. In particular, for $w < -1$ one has $c_s^2 <0$, which signals the presence of gradient instabilities. As shown in \cite{Creminelli:2006xe,Creminelli:2008wc} stability can be guaranteed by the presence of higher derivative operators but requires that the speed of sound is extremely small, practically zero \cite{Creminelli:2008wc}.

Regardless of the motivations expressed above on the stability of single field quintessence for $w<-1$, in the following we will be interested in considering the limit $c_s^2 \to 0$, which is obtained when $|\bar \rho_Q+\bar p_Q| \ll M^4$. We will see that what turns out to be physically relevant are the density and pressure perturbations on surfaces of constant $\phi$, i.e.~of constant $\pi$.  These are the perturbations in the so-called velocity orthogonal gauge, and using eq.~(\ref{deltarho_p}) they are given by
\be
\delta \rho^{\rm (v.o.)}_Q\equiv \delta \rho_Q- \dot{\bar \rho}_Q \pi = (\bar \rho_Q + \bar p_Q +4 M^4) \dot \pi \;, \qquad \delta p^{\rm (v.o.)}_Q\equiv \delta p_Q- \dot {\bar p}_Q \pi = (\bar \rho_Q +\bar  p_Q) \dot \pi \;.
\label{deltarho_vo}
\ee
Indeed, $c_s^2$ defined in eq.~(\ref{kessence2}) can be written as \cite{Creminelli:2008wc}
\be
c_s^2 = \frac{\delta p^{\rm (v.o.)}_Q}{  \delta \rho^{\rm (v.o.)}_Q} \;. \label{deltap}
\ee
Thus, the pressure perturbation is suppressed with respect to the energy density perturbation by the smallness of the speed of sound. As we will see, in the limit $c_s \to 0$ this implies that pressure forces are negligible and quintessence follows geodesics, remaining comoving with the dark matter. 

In the limit $c_s \to 0$, the energy density perturbation on velocity orthogonal slicing becomes 
\be
\delta \rho^{\rm (v.o.)}_Q = 4 M^4 \dot \pi\;.  \label{deltarhoQ}
\ee
Note that since $\dot \pi \sim H \pi$, the difference between $\delta \rho^{\rm (v.o.)}_Q$ and $\delta \rho_Q$ is negigible for small speed of sound, $\delta \rho^{\rm (v.o.)}_Q \simeq \delta \rho_Q $.
All these conclusions hold independently of the value of $\delta \rho_Q/\bar \rho_Q$, provided that the effective theory described by action (\ref{kessence2}) remains valid, i.e.~for $\dot \pi \ll 1$ \cite{ArkaniHamed:2005gu,Creminelli:2008wc}. In particular, they hold also when perturbations in the energy density of quintessence become non-linear, i.e., for $\delta \rho_Q \gg \bar \rho_Q$.

Gravitational perturbations can be straightforwardly included in the action (\ref{kessence2}) as in \cite{Creminelli:2008wc}. As a warm-up exercise we will here, instead, study
the evolution of quintessence in the spherical collapse solution.
According to the spherical collapse model, the overdensity can be described as a closed FLRW universe with a scale factor $R$ which is  different from the one of the background $a$. This remains true also when we take into account quintessence with negligible speed of sound. Indeed, eq.~(\ref{deltap}) shows that there is no pressure difference between the inside and the outside of the overdense region.
Therefore, inside the overdensity we can describe quintessence using eq.~(\ref{kessence2nd}), where the time evolution of the metric is described by the scale factor $R$ and we thus replace $a^3$ by $R^3$.

With this new action, the second term proportional to $\dot \pi$ can be integrated by parts and the coefficients of the linear part of the action rewritten in terms of $\bar \rho_Q$ and $\bar p_Q$ using eq.~(\ref{backrho_p}). However, now the linear part of the action does not cancel but can be written, using the background equation of motion, as $- \delta H (\bar \rho_Q+\bar p_Q)  \pi$, where we have defined $\delta H = H_{\rm in} - H$, with $H_{\rm in} \equiv \dot R/R$. We are thus left with a linear term in the action, due to the difference between the rates of expansion inside and outside the overdensity. After manipulations of the last two terms in eq.~(\ref{kessence2nd}), similarly to what was done to derive eq.~(\ref{kessence2}), the action inside the overdensity becomes\footnote{We will not include in the action the contribution to $\sqrt{-g}$ coming from the curvature of the closed FLRW universe. Indeed, as it is time independent, it does not affect our discussion.}
\begin{equation}
\begin{split}
  S =& \int {\rm d}^4x \; R^3 \Bigg[ \frac12 \left( \bar \rho_Q+\bar p_Q +4 M^4  \right) \dot{\pi}^2
  - \frac{1}{2}(\bar \rho_Q+\bar p_Q) \frac{(\vec \nab \pi)^2}{a^2}   + \frac{3}{2} \dot{H} (\bar \rho_Q+\bar p_Q)  \pi^2 \\
    &\qquad \qquad \quad - 3 \delta H (\bar \rho_Q + \bar p_Q) \pi
 - \frac{3}{2} \delta {H} (\bar \rho_Q+\bar p_Q)\dot{}  \; \pi^2 \Bigg] \, .
\end{split}
\label{eq:action}
\end{equation}

Using that $\dot \pi \sim H\pi$, neglecting time variations of $M$ and discarding terms suppressed when $| \bar \rho_Q + \bar p_Q| \ll M^4 $ (i.e., in the limit  $c_s^2 \to 0$) the equation of motion of $\pi$ derived from this action reads 
\begin{equation}
\ddot{\pi} + 3 \frac{\dot R}R  \dot{\pi} = - \frac{3 \delta H}{4M^4}(\bar \rho_Q + \bar p_Q)\,.
\label{pi_evolv_R}
\end{equation} 
As expected, the quintessence perturbation induced by the overdensity is proportional to $1+w$, i.e.~it vanishes in the limit of the cosmological constant. Note that the source term on the right-hand side of this equation can be written as $- 3 c_s^2 \delta H $ and is suppressed by the smallness of $c_s^2$. This implies that, even for large overdensities, i.e. $\delta H \gtrsim H$, variations of $\pi$ due to the gravitational potential well are extremely small, $\dot \pi \sim c_s^2$, inside the regime of validity of the effective theory. 
Furthermore, this also implies that the difference in $\pi$ between the homogeneous and closed FLRW solutions is also tiny, $\Delta \pi \sim c_s^2 H^{-1}$. Thus, the quintessential scalar field practically lies on the same point of its potential. 

Equation~(\ref{pi_evolv_R}) can be written, using eq.~(\ref{deltarhoQ}) (and $\delta \rho_Q \simeq \delta \rho^{\rm (v.o.)}_Q$), as
\begin{equation}
\label{deltarhoEOM}
\dot{\delta \rho}_Q + 3\frac{\dot R}R \delta \rho_Q = - {3 \delta H}{}(\bar \rho_Q + \bar p_Q)\,.
\end{equation} 
Note that, as $\delta H$ is always negative, the sign of $\delta \rho_Q$ is the same as that of $1+w$. Remarkably, this implies that for $w<-1$ dark matter halos accrete negative energy from quintessence, as was noticed at linear level in \cite{Weller:2003hw}.
Combining eq.~(\ref{deltarhoEOM}) with the background continuity equation, $\dot {\bar \rho}_Q + 3 H (\bar \rho_Q+\bar p_Q)=0$, we obtain 
\be
\dot{\rho}_Q + 3\frac{\dot R}R (\rho_Q + \bar p_Q) = 0 \,. \label{main_evolv}
\ee
This equation describes the evolution of the energy density of quintessence with $c_s^2=0$ inside a spherical overdensity dominated by dark matter. 
Note that the pressure perturbation is absent, as it is suppressed by $c_s^2 \to0$. Indeed, this equation differs from the description currently given in the literature for clustering dark energy. In particular, the analogue of this equation given in \cite{Mota:2004pa,Maor:2005hq} includes the pressure perturbation $\delta p_Q = w \delta \rho_Q$. Including the pressure perturbation $\delta p_Q$ leads to an incorrect description even in the linear regime, in contrast with eq.~(\ref{main_evolv}) which does match the linear theory for small overdensities.

In the following two sections we will make this analysis more complete and derive all the equations necessary to describe the spherical collapse with quintessence.

\section{Spherical collapse in local coordinates}
\label{sec:local_coords}

As we did in the former section, the spherical collapse is usually treated using FLRW coordinates, as in the simplest cases the overdensity evolves as an independent closed universe. This somewhat obscures a crucial simplification of the problem, i.e. that the collapse of dark matter haloes occurs on scales much smaller than
the Hubble radius.  In this limit one can treat gravity as a small perturbation of Minkowski space.\footnote{For a recent use of this approximation in cosmology see \cite{Nicolis:2008in}.} As the dynamics of $c_s=0$ quintessence is not completely intuitive, we want to make use of a coordinate system where this simplification is explicit;  this will also make the dynamics of the other cases of spherical collapse clearer.  We thus choose a coordinate system
around a given point, such that the deviation of the metric from Minkowski is suppressed by
$H^2 r^2$, where $r$ is the distance from the point, for any time.\footnote{In the spherically symmetric case, the range of validity of this approximation goes to zero close to the collapse singularity. However, this is not relevant because the
singularity is anyway an artifact of the spherical symmetry. In the
real case the curvature of space remains small and the halo reaches
virialization.} Notice we do not want to limit the validity of
our approximation to times shorter than $H^{-1}$ because this is also the typical
time-scale of the evolution of a dark matter halo. These requirements define the so called Fermi 
coordinates. 
Note also that we are not taking any Newtonian limit: as we are interested in quintessence we cannot neglect pressure as source of gravity.

A particular choice of Fermi coordinates are  the so-called 
Fermi normal coordinates  \cite{MM} 
where the deviation of the metric from Minkowski can be 
written as a Taylor expansion around the origin whose leading coefficients are components 
of the Riemann tensor. These are (with the convention of \cite{Misner:1974qy})
\begin{eqnarray}
g_{00} &=& -1 - R_{0l0m} |_{\vec 0} \; x^l x^m + \ldots\;, \\
g_{0i} &=& 0 -\frac23 R_{0lim} |_{\vec 0} \; x^l x^m+ \ldots\;, \\
g_{ij} &=& \delta_{ij} - \frac13 R_{iljm} |_{\vec 0} \; x^l x^m+ \ldots\;.
\end{eqnarray}
Here we will be interested only in spherically symmetric solutions. 
As $R_{0lim}$ vanishes because of rotational symmetry, $g_{0i}$ must be of order higher than $r^2$.
Thus we can neglect it in the following discussion. Furthermore, rotational symmetry implies 
that the corrections to $g_{00}$ will be proportional to $r^2$ while those to  
$g_{ij}$ will be proportional either to $r^2 $ or to 
$x_i x_j $. It is possible to make a redefinition of the radial coordinate such as to get rid of the term $x_i x_j $ without affecting $g_{00}$ and $g_{0i}$ at ${\cal O}(r^2)$. 
Note that in such a way we are using Fermi coordinates which are not of the normal form.
In this case the metric can be written in the Newtonian gauge (not to be confused with the cosmological perturbation theory Newtonian gauge) form as
\be
ds^2 = -(1+2\Phi) dt^2 + (1- 2\Psi) d \vec x^2\;, \label{Newtonian}
\ee
where $\Phi$ and $\Psi$ are proportional to $r^2$.
In this gauge the $00$ component of the Einstein equation gives
\be
\label{eq:nablapsi}
\nabla^2 \Psi = 4 \pi G \rho \;.
\ee
The part of the $ij$ Einstein equation proportional to the identity
gives
\be
6 \ddot\Psi + 2 \nabla^2 (\Phi - \Psi) = 24 \pi G p \;.
\ee
As the typical time scale is of order Hubble, the $\ddot \Psi$ term is suppressed with respect to $\nabla^2 \Psi$ by
${\cal{O}}(H^2 r^2)$ and can therefore be neglected. Thus, using
eq.~\eqref{eq:nablapsi} we obtain
\be
\nabla^2\Phi = 4\pi G(\rho + 3p) \;. 
\label{ein2}
\ee

As a first step, let us show how one can use these coordinates to describe an unperturbed FLRW solution with non-vanishing curvature. 
In isotropic comoving coordinates this metric is written as 
\be
ds^2 = - d\tau^2 + a(\tau)^2 \frac{d \vec y^{\,2}}{(1+\frac14 K \vec
  y^{\,2})^2} \;,
\ee 
where $K$ is the curvature parameter.
With the change of coordinates $\tau = t -\frac12 H r^2$ and $\vec y =
\frac{\vec x}{a} (1+ \frac14 H^2 r^2)$ \cite{Nicolis:2008in}, with $a$ and $H$ evaluated at $t$ rather than at $\tau$, one gets at first order in $H^2 r^2$,
\be
\label{eq:localg}
ds^2 \simeq - \left[1 -(\dot H + H^2) r^2 \right] dt^2 +  \left[1 -\frac12 (H^2
 + K/a^2 ) r^2 \right] d \vec x^{\,2} \;,
\ee
which is indeed of the Fermi form (\ref{Newtonian}), where the corrections from flat spacetime are given by
\be
\label{eq:phipsi}
\Phi = - \frac12
(\dot H +H^2) r^2\;, \qquad \Psi = \frac14 ( H^2 + K/a^2 ) r^2\;.
\ee

Let us now show that the metric (\ref{eq:localg}) is a solution of the Einstein equations (\ref{eq:nablapsi}) and (\ref{ein2}).
In the coordinates $(t,x^i)$, $\rho(\tau)$ and $p(\tau)$ are not space independent; however, their space dependence is suppressed by $H^2 r^2$ so that it can be neglected. 
With spherical symmetry, assuming regularity at the origin the two Einstein equations (\ref{eq:nablapsi}) and (\ref{ein2}) are then solved by
\be
\Psi = \frac{8 \pi G \rho}{3} \frac{r^2}{4} \;,
\ee
and 
\be
\label{eq:Phisol}
\Phi = \frac{4 \pi G}{3} (\rho + 3 p) \frac{r^2}{2} \;.
\ee
Comparing these expressions with  \eqref{eq:phipsi} we recover the two Friedmann equations, respectively, 
\be
H^2 + \frac{K}{a^2} = \frac{8 \pi G}{3}
\rho\;, 
\ee
and 
\be
\frac{\ddot a}{a} = - \frac{4 \pi G}{3} (\rho + 3p)\;.
\ee 
Note also that the
traceless part of the $ij$ Einstein equation, $(\partial_i \partial_j -
\frac13 \delta_{ij} \nabla^2)(\Phi-\Psi)=0$, is trivially satisfied by
the expressions above. Matter stays at fixed $\vec y$ in
the original FLRW coordinates; therefore it moves in the new
coordinates as $\vec x \propto a$, i.e.~with velocity $\vec v = H \vec
x$. Finally, using this equality one can check that also the $0i$ component of the
Einstein equation is satisfied.

Let us now look at the dynamical equations for the fluid. 
The time component of the conservation of the energy-momentum tensor gives (see for example \cite{WeinbergGC}) the continuity equation,
\be
\dot \rho + \vec \nabla \cdot\left[ (\rho +p) \vec v \right] =0\;. \label{continuity}
\ee
This equation is the same as in Minkowski spacetime as the gravitational corrections only induce terms 
suppressed by ${\cal O}(H^2 r^2)$.
When the velocity $\vec v$ 
is simply given by an unperturbed Hubble flow we obtain the standard
conservation equation in expanding space, $\dot \rho + 3H (\rho+p) =0$.

The spatial component of the conservation of the energy-momentum tensor gives the Euler equation,
\be
\dot {\vec v} + (\vec v \cdot \vec \nabla) \vec v = - \frac{1}{ (\rho + p)} \left[\vec \nabla p + \vec v \;\frac{\partial p}{\partial t} \right]  - \vec \nabla \Phi
\;, \label{Euler} 
\ee
where we have assumed that $v \ll c$. At leading order in ${\cal O}(H^2 r^2)$ gravitational perturbations enter only through the last term on the right-hand side of this equation. 
In the particular case of an isotropic and homogenous solution
the first term on the right-hand side exactly cancels: 
as $\vec \nabla p = - \dot p H \vec x$, the gradient of the pressure cancels with the term coming from its time dependence. 
This is not surprising as what matters in the Euler equation is the 4-dimensional 
gradient of pressure perpendicular to the fluid 4-velocity. 
In this case eq.~(\ref{Euler}) reduces to
\be
\dot {\vec v} + (\vec v \cdot \vec \nabla) \vec v = - \vec \nabla \Phi
\;. \label{Euler2}
\ee
This equation is verified by the Hubble flow $\vec v = H \vec x$ since we get
\be
\label{eq:EulerFRW}
\frac{\ddot a }{a} \vec x = - \vec\nabla\Phi \;,
\ee
which is clearly satisfied by the explicit expression for $\Phi$, eq.~\eqref{eq:phipsi}.

We can now use these local coordinates to describe the spherical collapse in various cases, 
starting from the simplest.


\subsubsection*{$\bullet \ $ Dark matter only}

\begin{figure}[!!!h]
\begin{center}
\includegraphics[scale=0.8]{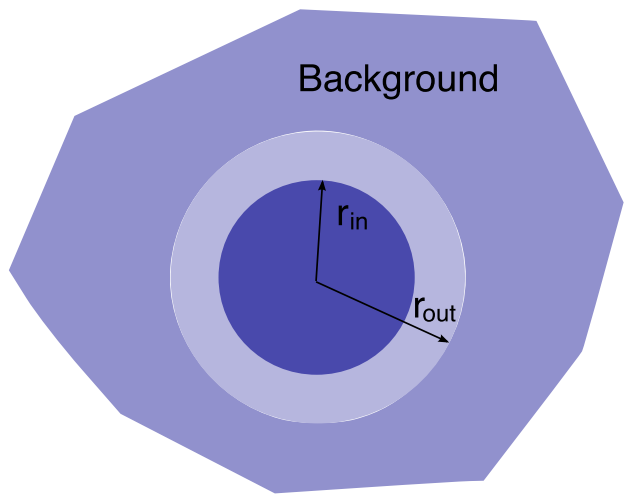}  
\end{center}
\caption{\small {\em Spherical collapse}}
\label{fig:spheres}
\end{figure}

Let us take a spherically symmetric distribution of matter around the origin. As both the gravitational potentials $\Phi$ and $\Psi$ satisfy the Poisson equation, we do not need to know how the mass is radially distributed to solve for the gravitational background outside a given radius $r$.  We just need the total mass inside the radius $r$.
In particular (see figure \ref{fig:spheres}) if inside a given radius $r_{\rm out}$, a distribution contains as much matter as the unperturbed cosmological solution, from the outside it will look exactly as the unperturbed background.
This implies that we can smoothly glue this solution at $r = r_{\rm out}$ to the cosmological background, and that the latter will not be affected by the gravitational collapse inside. This is of course a linearized version of Birkhoff's theorem in General Relativity.

Conversely, the solution inside a given radius is not affected by what happens outside. In particular, if we assume a homogeneous initial condition inside a radius $r_{\rm in}$ (with $r_{\rm in} < r_{\rm out}$), this central region will evolve as if these homogeneous initial conditions were extended outside, i.e.~as a complete FLRW solution \cite{Gunn:1972sv}. 
The central overdense region will remain exactly homogeneous, reaching maximum expansion and then collapsing.

Without further assumptions, the evolution of the layer $r_{\rm in} < r < r_{\rm out}$ does not enjoy particular simplifications and its evolution must be computed as a function of the initial profile. In any case, this is usually irrelevant as we are only interested in the fate of the $r < r_{\rm in}$ region. If we assume that the layer  $r_{\rm in} < r < r_{\rm out}$ is empty, the Poisson equations \eqref{eq:nablapsi} and \eqref{ein2} give $1/r$ solutions for the potentials, like for a source localized at the origin. This is the linearization of the exact Schwarzschild solution.


\subsubsection*{$\bullet \ $ Dark matter and a cosmological constant }

The considerations above also hold when we include a cosmological constant. Although $\Phi$ and $\Psi$ now solve different equations (because $p_\Lambda \neq 0$), they are both of the Poisson form. Thus, we still have an unperturbed evolution outside $r_{\rm out}$ if the total matter inside matches the background value. Assuming initial homogeneity, the central region $r < r_{\rm in}$ will evolve like a complete FLRW universe \cite{Lahav:1991wc}. Although now pressure does not vanish, it just comes from the cosmological constant which does not define a preferred frame and is therefore comoving with dark matter both inside and outside the overdensity.


\subsubsection*{$\bullet \ $ Non-clustering quintessence: ${\mathbf c_s=1}$ } 
When quintessence has a speed of sound $c_s=1$, it  does not effectively cluster but it keeps on following the cosmological background solution, irrespective of the dark matter clustering \cite{Wang:1998gt}.

As before, outside $r_{\rm out}$ there is an unperturbed cosmological background. What is new now is that the central region $r < r_{\rm in}$ does not behave as a complete FLRW solution, even if we start with a homogeneous overdensity. Indeed, quintessence and dark matter do not have a common velocity: while dark matter slows down and eventually starts collapsing, quintessence keeps following the external Hubble flow $\vec v_{\rm Q} = H_{\rm out} \vec x$. Note that, on the other hand, in the cosmological constant case one cannot define a dark energy 4-velocity as its energy-momentum tensor is proportional to the metric.
To study the evolution of the dark matter overdensity one must use the Euler equation \eqref{eq:EulerFRW}. 
Here, what defines the velocity flow of dark matter is the effective ``scale factor'' $R$, $\vec v_\m = \dot R/R \; \vec x$. This yields
\be
\frac{\ddot R}{R} \vec x = - \vec\nabla\Phi \;.
\ee
Using the explicit solution \eqref{eq:Phisol} for $\Phi$, this equation becomes \cite{Wang:1998gt}
\be
\label{sndfried}
\frac{\ddot R}{R} = -\frac{4 \pi G}{3} (\rho_\m + \bar\rho_{\rm Q} + 3 \bar p_{\rm Q}) \;,
\ee
where we have separated the contribution of dark matter and quintessence to $\Phi$. Notice that, although this equation looks like one of the Friedmann equations, the dynamics of $R$ is not the same as for a FLRW universe. Indeed, $\rho_\m$ evolves following the scale factor $R$, while the quintessence follows the external scale factor $a$. In a FLRW universe, from eq.~\eqref{sndfried} together with the continuity equation one can derive the first Friedmann equation, $(\dot R/R)^2 = \frac{8 \pi G}{3} \rho - K/R^2$. Here, as the different components follow different scale factors, this is not longer possible and the first Friedmann equation does not hold.


\subsubsection*{$\bullet \ $  Clustering quintessence: ${\mathbf c_s=0}$ }
Let us now move to the subject of this paper. 
We want to show that in the limit of vanishing speed of sound quintessence remains comoving everywhere with dark matter.  In particular, this implies that in the region $r<r_{\rm in}$ quintessence follows dark matter in the collapse and the overdensity behaves as an exact FLRW solution so that, contrary to the $c_s=1$ case, also the first  Friedmann equation holds. The fact that quintessence remains comoving with dark matter can be understood both by using the fluid equations or directly from the scalar field equation of motion.

In the fluid language the dynamics is described by the Euler equation (\ref{Euler}). In general, in the presence of sizable pressure gradients a fluid does not remain comoving with dark matter, i.e.~it does not follow geodesics. Since quintessence has a sizable pressure, the fact that it moves following geodesics may be unexpected but it is obtained in the limit $c_s \to 0$. This can be easily seen by rewriting the Euler equation for quintessence in covariant form as
\be
 u^\mu \nabla_\mu u^\nu = - \frac1{(\rho_Q+p_Q)} (g^{\nu \sigma} + u^\nu u^\sigma) \nabla_\sigma p_Q\,,
\label{geodesic}
\ee
where $u^\mu$ is the quintessence 4-velocity. When the right-hand side of this equation vanishes,  the 4-velocity solves the geodesic equation. 
Notice that the pressure gradient is multiplied by the projector perpendicular to the fluid 4-velocity. This is the same as projecting on surfaces of constant $\phi$ and it is equivalent to a gradient of the velocity-orthogonal pressure perturbation that appears in equation (\ref{deltarho_vo}), which involves only $\dot \pi$, and not $\pi$.
By eq.~(\ref{deltap}) this is negligible in the limit $c_s \to 0$ and thus the right-hand side of (\ref{geodesic}) vanishes.

This result is even clearer in the scalar field language. Taking the derivative of the equation defining the quantity $X$ in \eqref{kaction},
\be
\partial^\nu(\partial^\mu\phi \partial_\mu\phi) = - \partial^\nu X \;,
\ee 
and writing it in terms of the 4-velocity $u^\mu = - \partial^\mu\phi/\sqrt{X}$ we have 
\be
2 u^\mu \sqrt{X} \nabla_\mu (\sqrt{X} u^\nu) = - \partial^\nu X \;,
\ee
and therefore
\be
u^\mu \nabla_\mu u^\nu = -\frac1{2 X} (g^{\nu \sigma} + u^\nu u^\sigma) \partial_\sigma X \;.
\ee
Equation \eqref{geodesic} is recovered using eq.~\eqref{rho_P_u} and taking into account that $\partial_\sigma P(\phi,X) = \partial P/\partial\phi \cdot \partial_\sigma \phi + \partial P/\partial X \cdot \partial_\sigma X$ and that the first term vanishes when multiplied by the projector orthogonal to $u^\mu$. From this we clearly see that what matters is only the gradient of the pressure on $\phi=$ const hypersufaces. This vanishes in the limit $c_s \to 0$ and thus we have geodesic motion. We stress that, although quintessence with $c_s = 0$ follows geodesics, its dynamics is quite different from dark matter. Pressure does not accelerate the quintessence 4-velocity but it does affect the energy conservation equation \eqref{continuity}. Moreover, quintessence does not enjoy a conserved current, while dark matter particle number is conserved; this is related to the absence of the shift symmetry $\phi \to \phi + c$ in the scalar field Lagrangian (see for example \cite{Boubekeur:2008kn}).

As discussed in section \ref{sec:model}, the different dark matter evolution inside and outside the overdensity changes the quintessence solution by  a very tiny amount $\Delta\pi \sim c_s^2 H^{-1}$: the quintessence field sits at the same position along its potential, $\phi =\bar\phi(t)$, apart from negligible $c_s^2$ corrections. Notice that this was derived using two different Friedmann coordinate systems, one following dark matter inside the overdensity and one following the unperturbed Hubble flow outside. Thus, in reality we have two solutions $\phi=\bar\phi(t_{\rm in})$ and $\phi=\bar\phi(t_{\rm out})$ respectively.
Once these two solutions are written in the same local coordinates \eqref{eq:localg}, the solution for $\phi$ becomes $ \phi = \bar \phi (t - \frac12 H_{\rm in} r^2)$ for $r < r_{\rm in}$ and $ \phi = \bar \phi (t - \frac12 H_{\rm out} r^2)$ for $r > r_{\rm out}$. This implies that in these coordinates $\pi $ has to jump in the layer between the two regions, by an amount $\Delta \pi \sim \frac{1}{2} r^2 \delta H $, and that this jump is not suppressed by $c_s^2$. One may expect that the scalar field would ``react" to this gradient between the inside and the outside. However, this does not happen in the limit $c_s \to 0$ as the spatial kinetic term is very suppressed. Let us see this explicitly. 

To study the scalar field equations in the local coordinates, one can start by writing the equations in Minkowski space and then check a posteriori that the deviation of the metric from flat space only gives relative corrections ${\cal{O}}(H^2 r^2)$. 
The evolution equation for $\phi$, eq.~(\ref{phi_evolution}), reads in Minkowski space
\be
- \partial_t (P_{,X} \dot \phi) + \partial_i (P_{,X} \partial_i \phi) = -\frac12 P_{,\phi} \;.
\ee
If we try a comoving solution of the form $\phi = \bar \phi(t -\frac12 H r^2)$ we end up with the standard FLRW equation for $\bar \phi$, inclusive of the friction term,
\be
- \partial_t (\bar P_{,X} \dot{\bar \phi}) - 3 H  \bar P_{,X} \dot{\bar \phi}
= -\frac12 \bar P_{,\phi} \;.
\ee
Metric fluctuations give only a correction to this equation of order ${\cal O} (H^2 r^2)$.
As we discussed, the two homogeneous solutions for $r < r_{\rm in}$ and $r > r_{\rm out}$ are different so that we expect gradient terms to smooth out the initial top-hat profile. To estimate the thickness $L$ of the layer over which the smoothing takes place, we can study perturbations around a top-hat profile and require the spatial and time kinetic term of the perturbation $\pi$ in eq.~(\ref{kessence2}) to be comparable,
\be
M^4 \dot \pi^2 \sim (\rho_Q+p_Q) \frac{\pi^2}{L^2} \;.
\ee
Using that $\dot \pi \sim H \pi$,  from this comparison we obtain $L \sim |c_s| H^{-1}$. This makes perfect sense: our top-hat profiles are smoothed out over a distance comparable to the sound horizon.\footnote{It is straightforward to check that this estimate is not altered by the higher derivative operators that are required for stability when $w<-1$ \cite{Creminelli:2006xe}.}

In conclusion, the solutions outside and inside the overdensity are exact FLRW with quintessence comoving with dark matter. Gradient terms will smooth out this solutions on scales of order of the sound horizon, which vanishes for $c_s \to 0$.
This discussion also tells us that taking $c_s =0$ will be correct only for objects which are much bigger than the sound horizon $|c_s| H^{-1}$. In the opposite limit of an object which is much smaller than the sound horizon, one can treat quintessence as unperturbed as discussed above in the $c_s^2 =1$ case. For example, if one is interested in objects larger than $1$ Mpc, one can neglect the speed of sound as long as $|c_s| \lesssim 10^{-4}$.


\section{Solving the spherical collapse}
\label{sec:solving}

In this section we derive the equations for the spherical collapse of dark matter in the presence 
of quintessence with vanishing speed of sound and we compute their solutions numerically.


\subsubsection*{$\bullet \ $   The background universe} The background is described by a flat FLRW metric with scale factor satisfying the Friedmann equation,
\begin{equation}
\left(\frac{\dot{a}}{a}\right)^2 = \frac{8\pi G}{3}(\bar{\rho}_m + \bar{\rho}_Q)\,, \label{Feq}
\end{equation}
where $\bar{\rho}_m$ and $\bar{\rho}_Q$ are the background energy density of dark 
matter and quintessence, respectively.
For later purposes, we express $\bar{\rho}_m$ and $\bar{\rho}_Q$ in terms of the
fractional abundance of dark matter $\Om_m$,
\be
  \bar{\rho}_m  \equiv \frac{3 H^2}{8 \pi G} \Om_m\,,\qquad
  \bar{\rho}_Q = \frac{1 - \Om_m}{\Om_m} \bar{\rho}_m\,. \label{rhomQ}
\ee
Dark matter redshifts with the expansion as the physical volume, $\bar{\rho}_m \propto a^{-3}$, 
while the energy density of quintessence scales as $\bar{\rho}_Q \propto a^{-3(1+w)}$. The dark matter contribution to the critical density $\Om_m$ can be written as a function of its value today, $\Om_{m,0}$, and $x$, the scale factor normalized to unity today (at $t=t_0$), 
\be
x \equiv a/a_0\; .  \label{def_x}
\ee
This yields
\begin{equation}
  \Om_m(x) = \left( 1 + \frac{1 - \Om_{m,0}}{\Om_{m,0}} x^{-3 w} \right)^{-1}\,. \label{Omegax}
\end{equation}
Equation~(\ref{rhomQ}) can be then rewritten as
\be
\bar{\rho}_m  
  = \frac{3 H^2_0}{8 \pi G} \frac{\Om_{m,0}}{x^3} \,,\qquad
  \bar{\rho}_Q = \frac{1 - \Om_{m,0}}{\Om_{m,0}} x^{-3 w} \bar{\rho}_m\,, \label{useful_rel1}
\ee
where the second equation follows from (\ref{Omegax}).
Furthermore, rescaling the time variable by defining 
\be
\eta \equiv \sqrt{\Omega_{m,0}} \; H_0  t\;, \label{def_eta}
\ee
one can rewrite the Friedmann equation as
\begin{equation}
\frac{{\rm d}x}{{\rm d}\eta} = (x\Omega_m(x))^{-1/2}\,. \label{dxdt}
\end{equation}
The initial condition for $x$ can be imposed at some small initial time $\eta_i$ during  matter dominance, 
$x_i = (3\eta_i/2)^{2/3}$. Then, eqs.~(\ref{Omegax}) and (\ref{dxdt}) completely describe the background evolution of the metric and energy-momentum tensors.


\subsubsection*{$\bullet \ $   The linear evolution } Before studying the collapsing spherical overdensity we derive the evolution equations 
of perturbations of dark matter and quintessence in the linear regime. As we consider scales much smaller than the Hubble radius, the gauge dependence is not important. We will thus perturb 
the continuity and Euler equations in local coordinates, eqs.~(\ref{continuity}) and (\ref{Euler2}), 
adding small inhomogeneous perturbations $\delta(t,\vec x)$ and $\vec u(t,\vec x)$ to the homogeneous energy density and Hubble flow velocity, 
\be
\rho  = \bar \rho (1 + \delta )\;, \qquad \vec v  = H \vec x + \vec u\;.
\ee
Let us start from the dark matter. Perturbing at linear order eq.~(\ref{continuity}) with $p_m=0$ yields
\be
\left( \frac{\partial}{\partial t}+ H \vec x \cdot \vec \nabla \right) \delta_m =  - \vec \nabla \cdot
\vec u  \; \qquad {\rm (local \ coords})\;,
\ee
where we have specified that we are describing perturbations using local spatial coordinates $\vec x$. On the other hand, on the left-hand side of this equation one recognizes the time derivative at fixed comoving 
coordinates $\vec y = \vec x /a(t)$, i.e.,
\be
\left( \frac{\partial}{\partial t}\right)_{\vec y} = \left( \frac{\partial}{\partial t}+ H \vec x \cdot \vec \nabla \right)_{\vec x} \;.
\ee
Indeed, here we are interested in describing the evolution of an overdensity of dark matter 
contained in a comoving volume. Thus, we describe $\delta_m$ and $\vec u$ 
as a function of the comoving coordinates, which simply gives
\be
\dot \delta_m + \frac{1}{a} \vec \nabla \cdot \vec u =0 \;. \label{delta_linear}
\ee

To close this equation we need the evolution of the dark matter peculiar velocity $\vec u$. This can be obtained by 
perturbing at linear order the Euler equation (\ref{Euler2}). Using comoving coordinates
the perturbed Euler equation becomes
\be
\dot{\vec u} + H \vec u + \frac{1}{a} \vec \nabla \delta \Phi=0\;, \label{u_linear}
\ee 
where $\delta \Phi$ is the perturbation of the Newtonian potential,
\be
\delta \Phi =\Phi + \frac{1}{2}(\dot H + H^2) r^2\;. \label{phi}
\ee
Equations (\ref{delta_linear})--(\ref{phi}) have been derived for instance in \cite{Peebles:1994xt} in the context of Newtonian mechanics described with expanding coordinates, for a pressureless fluid in the presence of vacuum energy.
Here the Poisson equation for $\delta \Phi$ is  sourced by both dark matter and quintessence perturbations,
\be
\frac1{a^2}\nabla^2 \delta \Phi = 4 \pi G (\bar \rho_m \delta_m + \bar \rho_Q \delta_Q)\;,\label{phi2}
\ee
where we have used that $\delta p_Q=0$.
The final step is to eliminate the peculiar velocity by subtracting the divergence of eq.~(\ref{u_linear}) from the time derivative of eq.~(\ref{delta_linear}). With the Poisson equation (\ref{phi2}) we obtain
\begin{equation}
\ddot{\delta}_m + 2H\dot{\delta}_m = 4\pi G( \bar{\rho}_m \delta_m+ \bar{\rho}_Q \delta_Q ) \,.
\label{deltam_linear}
\end{equation}

For quintessence we perturb the continuity equation (\ref{continuity}) which gives, in comoving coordinates, using $\delta p_Q=0$,
\be
\dot \delta_Q - 3 H w \delta_Q + (1 + w) \frac{1}{a} \vec \nabla \cdot \vec u  =0 \;. \label{deltaQ_linear}
\ee
To eliminate the divergence of the peculiar velocity we can use eq.~(\ref{delta_linear}) taking quintessence to be comoving with dark matter. Indeed, as explained above, both the dark matter and quintessence follow geodesics and are dragged by the same potential well and the growing mode of their velocities is the same. Thus
\be
\dot \delta_Q - 3 H w \delta_Q  = (1+w) \dot \delta_m \;. \label{deltaQ_linear2}
\ee
In matter dominance, when $\delta_m \propto a$, the solution of this equation is \cite{Creminelli:2008wc}
\be
\delta_Q = \frac{1+w}{1-3 w} \delta_m\;. \label{deltaqm}
\ee
Note that the denominator on the right-hand side further suppresses the perturbation of quintessence with respect to the naive $1+w$ estimate.

In terms of the dimensionless variables $x$ and $\eta$, respectively defined in eqs.~(\ref{def_x}) and (\ref{def_eta}), equations \eqref{deltam_linear} and \eqref{deltaQ_linear2} rewritten as
\begin{equation}
\frac{{\rm d}^2\delta_m}{{\rm d}\eta^2} + \frac{2}{x}\frac{{\rm d} x}{{\rm d}\eta} \frac{{\rm d}\delta_m }{{\rm d} \eta} = \frac{3}{2 x^3}\left(\delta_m +  \frac{1 - \Omega_{m,0}}{\Omega_{m,0}} x^{-3w} \delta_Q  \right) \,, \label{master4}
\end{equation}
where we have used eq.~(\ref{useful_rel1}), and
\begin{equation}
\frac{{\rm d}\delta_Q}{{\rm d}\eta} - \frac{3}{x}\frac{{\rm d}x}{{\rm d}\eta}w \delta_Q = (1+w)\frac{{\rm d}\delta_m}{{\rm d}\eta}\,. \label{master5}
\end{equation}
The initial conditions are set in terms of the initial dark matter density contrast $\delta_{m, i}$. In matter dominance $\dot \delta_m = H\delta_m$ i.e., ${{\rm d}}\delta_m/{{\rm d}\eta}|_{i} = 2 \delta_{m,i}/(3\eta_i)$, while the value of $\delta_{Q,i}$ is fixed by $\delta_{m,i}$ through equation (\ref{deltaqm}).


\subsubsection*{$\bullet \ $   The spherical overdensity } 
We now study the evolution of a spherical homogeneous overdensity of radius $R$ in a FLRW  background that satisfies the Friedmann equation (\ref{Feq}). We denote the energy densities of dark matter and quintessence inside the collapsing ball by $\rho_m$ and $\rho_Q$, respectively. Since dark matter is pressureless $p_m =0$ and since quintessence pressure perturbation is negligible, $\delta p_Q \ll \delta \rho_Q$, we can take quintessence pressure to be the unperturbed one $\bar p_Q $.

In local coordinates, the evolution of the scale factor $R$ is described by the Euler equation (\ref{eq:EulerFRW}). Using the appropriate scale factor -- i.e., $R$ instead of $a$ -- and replacing the potential $\Phi$ using eq.~(\ref{eq:Phisol}), the divergence of this equation can be written as
\begin{equation}
  \frac{\ddot{R}}{R} = - \frac{4 \pi G}{3} \left( \rho_m + \rho_Q
    + 3 \bar{p}_Q \right) \, . \label{R_evol}
    \end{equation}
(Note that for a non-clustering quintessence the equation for $R$ is the same with $\rho_Q$ replaced by
$\bar{\rho}_Q$~\cite{Wang:1998gt}.)

For the evolution equations for $\rho_m$ and $\rho_Q$ we use the continuity equation (\ref{continuity}). Inside the ball this reads, for dark matter,
\begin{equation}
  \dot{\rho}_m + 3 \frac{\dot{R}}{R} \rho_m = 0 \, ,
\end{equation}
whose solution is simply
\begin{equation}
  \rho_m = \rho_{m,i} \frac{R_i^3}{R^3}\;. \label{rhom_sol}
\end{equation}
For dark energy eq.~(\ref{continuity}) becomes 
\begin{equation}
  \dot{\rho}_Q + 3 \frac{\dot{R}}{R} (\rho_Q + \bar{p}_Q)= 0\;, \label{rhoQ_evol_in}
\end{equation}
which can be rewritten in terms of the nonlinear density contrast $\Delta_Q \equiv \rho_Q/\bar \rho_Q-1$  as
\begin{equation}
   \dot{\Delta}_Q + 3 \frac{\dot{R}}{R} \Delta_Q
   - 3 \frac{\dot{a}}{a} (1 + w) \Delta_Q
   + 3 (1 + w) \left( \frac{\dot{R}}{R} - \frac{\dot{a}}{a} \right) = 0 \, . \label{DeltaQ_evol}
 \end{equation}

To solve eqs.~(\ref{R_evol}) and (\ref{DeltaQ_evol}) numerically it is convenient to use $y$, the radius of the ball normalized to unity at the initial time,
\be
y \equiv R/R_i\;,
\ee
and change $a$ and $t$ to the dimensionless variables $x$, $\eta$. Using eq.~(\ref{useful_rel1}), eq.~(\ref{R_evol}) can be rewritten as
\be
  \frac{\rmd^2 y}{\rmd \eta^2}
  + \frac{1}{2} \left[ \frac{1+\del_{m, i}}{x_i^3} \frac{1}{y^2}
    + (1 + 3w + \Delta_Q) \frac{1 - \Om_{m,o}}{\Om_{m,o}} \frac{y}{x^{3(1+w)}} \right]=0\,, \label{master2}
\ee
where we have used eq.~(\ref{rhom_sol}) and that in the linear regime, where the initial conditions are set, $\rho_{m}/\bar \rho_{m}|_i =1+ \delta_{m,i}$. Equation (\ref{DeltaQ_evol}) yields
\be    
  \frac{\rmd \Delta_Q}{\rmd \eta}
  + 3 (1 + w) \left( \frac{1}{y} \frac{\rmd y}{\rmd \eta}
    - \frac{1 + \Delta_Q}{x} \frac{\rmd x}{\rmd \eta} \right)
  + 3 \frac{\Delta_Q}{y} \frac{\rmd y}{\rmd \eta}=0 \,. \label{master3}
\ee
As initial conditions we have $y_i = 1$ by definition; the expansion rate of a collapsing sphere with dark matter only and in the linear regime can be written as \cite{Liddle:2000cg} 
\be
\frac{\dot R}{R} = \frac{2}{3t} \left(1 - \frac{1}{3} \delta_m \right)\,,
\ee
which fixes the first derivative of $y$, ${{\rm d} y}/{{\rm d}\eta}|_i = {2} (1-\delta_{m,i}/3)/({3 \eta_i})$. For the dark energy perturbation we use that $\Delta_{Q,i}$ is linear at early times, $\Delta_{Q,i} = \delta_{Q,i}$, and thus is fixed in terms of $\delta_{m,i}$ by eq.~(\ref{deltaqm}).

\begin{figure}[!!!h]
\begin{center}
\includegraphics[scale=1.2]{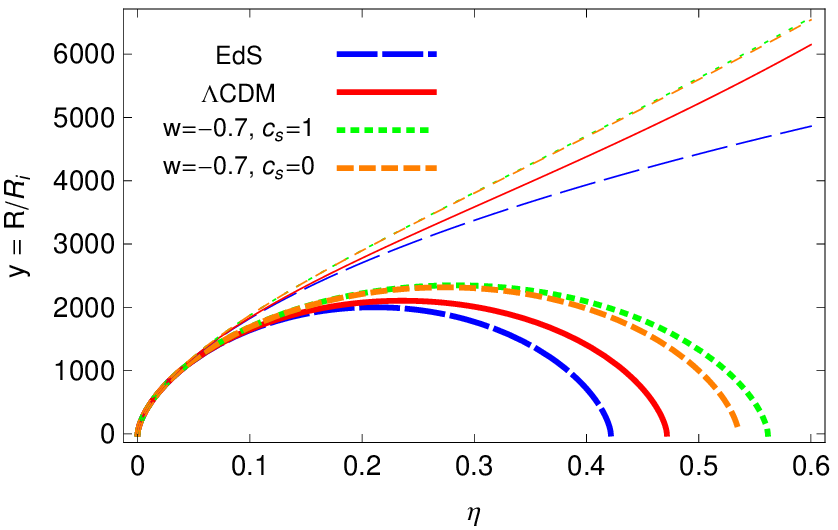}  

\vspace{0.5cm}
\includegraphics[scale=1.2]{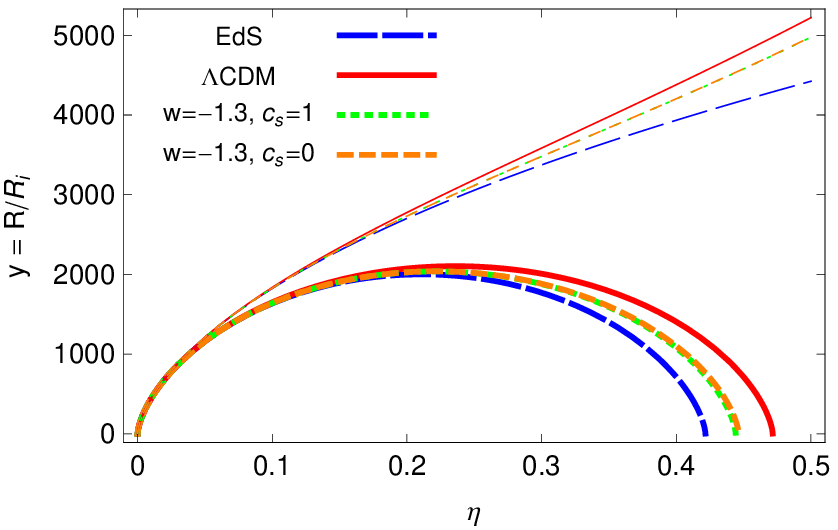}
\end{center}
\caption{\small {\em Thick lines: time evolution of the radius for a spherical collapse. Thin lines: time evolution following the linearized solutions. The quintessence equation of state is $w=-0.7$ (above) and $w=-1.3$ (below). Starting with the same overdensity, a model with CDM only is the first to collapse. In the upper figure $\Lambda$CDM collapses before the quintessence models as dark energy with $w=-0.7$ is more important in the past. The situation is reversed for $w=-1.3$. For $w=-0.7$ the $c_s=0$ quintessence collapses before $c_s=1$ as positive energy clusters together with dark matter. For $w=-1.3$ the situation is reversed as negative energy clusters and hinders the collapse. Note that quintessence models with $c_s^2=1$ and $w<-1$ are plagued by ghost instabilities and are thus very pathological on short scales. In this figure and in the following ones we study this case only for comparison with the $c_s^2 =0$ case.}}
\label{fig:collapse}
\end{figure}

By solving numerically eq.~(\ref{dxdt}) for the background evolution described by $x$ and plugging the result into the coupled eqs.~(\ref{master2}) and (\ref{master3}), one can compute the evolution of $R$ as a function of time $t$, from the initial time $t_i$ to the time of collapse $t_c$. 
The evolution of $R$ is shown in figure~\ref{fig:collapse} for four different models: CDM only, $\Lambda$CDM, $c_s=1$ quintessence and $c_s=0$ quintessence in the cases $w=-0.7$ and $w=-1.3$. We stress that quintessence models with $c_s^2=1$ and $w<-1$ are plagued by ghost instabilities and are thus very pathological on short scales. We study them here only for comparison with the $c_s^2 =0$ case. We asume
 $\Omega_{m,0}$ and $\Omega_{Q,0}$  (or $\Omega_{\Lambda}$) to be the WMAP5 best fit values \cite{Komatsu:2008hk}.  We have taken $\delta_{m,i}=3 \cdot 10^{-4}$ as initial dark matter overdensity at $\eta_i=10^{-6}$. 
As expected, since the cosmological constant and the quintessence slow down the evolution of $R$, the collapse is faster in the pure CDM model. This effect takes place earlier for $w > -1$ as in this case quintessence is more important in the past than the cosmological constant. Thus, for $w=-0.7$ the collapse happens later. On the contrary, for a quintessence with $w=-1.3$ collapse takes place earlier. For $w=-0.7$ the collapse is enhanced by the quintessence perturbations and it takes place faster when $c_s^2=0$. The opposite happens for $w<-1$, as in this case negative energy clusters, hindering the collapse (see eq.~(\ref{deltarhoEOM})).

\begin{figure}[!!!h]
\begin{center}
\includegraphics[scale=1.2]{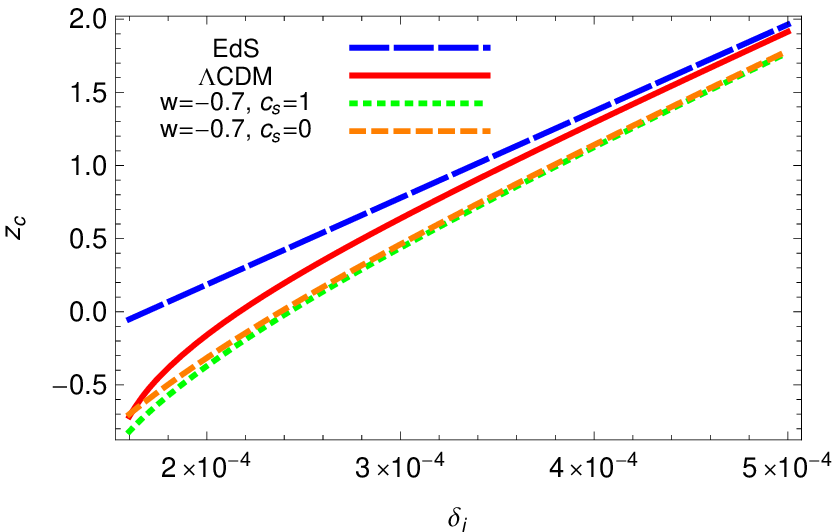}  

\vspace{0.5cm}
\includegraphics[scale=1.2]{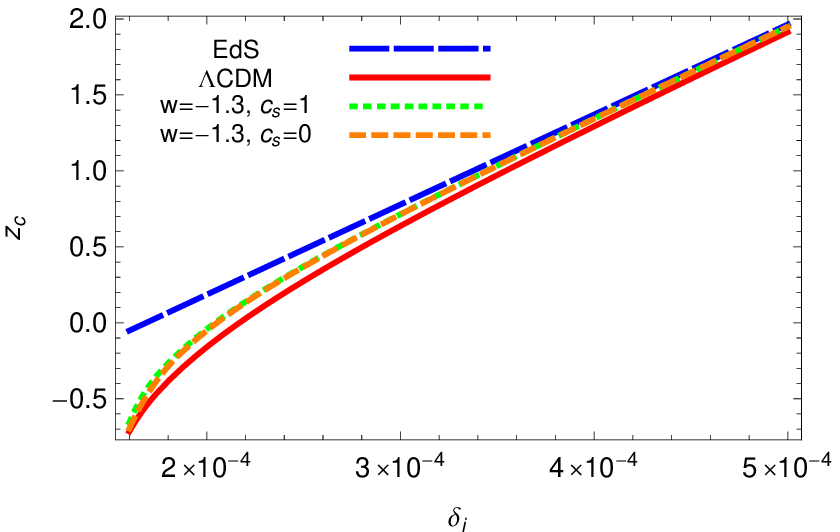}
\end{center}
\caption{\small {\em  Redshift of collapse as a function of the initial overdensity. In the upper figure quintessence models have $w=-0.7$, while they have $w=-1.3$ in the lower one. The behavior follows that explained in figure~\ref{fig:collapse}.}}
\label{fig2}
\end{figure}

In general, the time of collapse depends on the value of the initial dark matter overdensity. This is shown in figure~\ref{fig2}, where the redshift of collapse $z_c$ is plotted as a function of the initial density constrast $\delta_{m,i}$ at the same initial time. As expected, larger ovedensities collapse earlier, at higher redshift. For large enough overdensities -- and early enough collapse -- the redshift of collapse becomes the same for all four different models, because the cosmological constant or the quintessence remain subdominant during the whole process.
As expected, quintessence with $w>-1$ requires a larger initial overdensity to collapse and in this case quintessence perturbations ($c_s^2=0$) help the collapse. The opposite happens for $w<-1$.

\begin{figure}[!!!h]
\begin{center}
\includegraphics[scale=1.2]{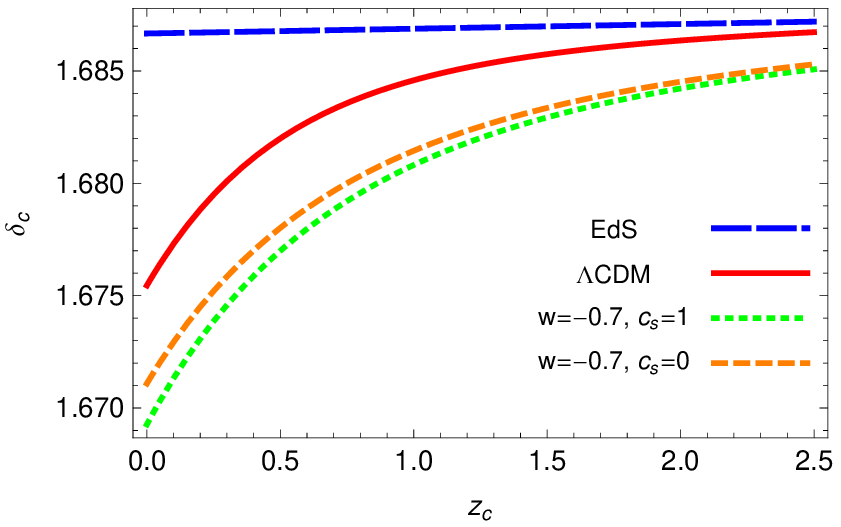}  

\vspace{0.5cm}
\includegraphics[scale=1.2]{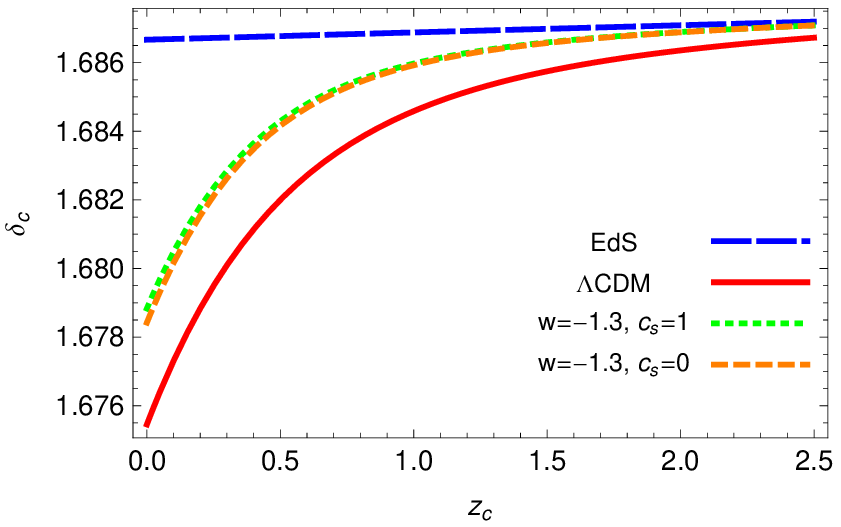}
\end{center}
\caption{\small {\em Linear overdensity at collapse as a function of the redshift of collapse.  In the upper figure the quintessence models have $w=-0.7$, while they have $w=-1.3$ in the lower one.}}
\label{threshold}
\end{figure}

An important quantity to compute in order to derive the mass function is the critical density contrast $\delta_c$, i.e., the density contrast in the linear theory computed at the time when the spherical collapse solution reaches the singularity. Thus, we numerically solve the linear evolution equations for $\delta_m$ and $\delta_Q$, eqs.~(\ref{master4}) and (\ref{master5}), and we take $\delta_c$ to be $\delta_m$ at the time of collapse. In the standard CDM scenario $\delta_c$ is given by the well-known number $1.686$ \cite{Gunn:1972sv} independently of the redshift of collapse $z_c$. However, in the presence of a cosmological constant or quintessence, $\delta_c$ depends on the redshift of collapse. Indeed, as the relative abundance of dark matter and dark energy changes with time, the dynamics of the spherical collapse depends on when it takes place. This is shown in figure~\ref{threshold}, where we plot $\delta_c$ as a function of $z_c$. This result generalizes to quintessence with $c_s^2=0$ the standard results obtained for CDM \cite{Gunn:1972sv}, $\Lambda$CDM \cite{Lahav:1991wc} and smooth quintessence \cite{Zlatev:1998tr}. 
As expected, if the collapse takes place early, when the cosmological constant or quintessence are not important, the critical density $\del_c$ will be the same as for CDM. The cosmological constant decreases the value of $\del_c$ and quintessence with $w > -1$, becoming important earlier, decreases it even more. Quintessence perturbations enhance $\del_c$ if $w > -1$. This effect is very mild for $w <-1$ because quintessence becomes important only at very late time.

The change of threshold is very small ($\lesssim 0.5 \%$)
\cite{Percival:2005vm} 
in all the cases and it is easy to understand why. Let us compare for example a universe with CDM only with a $\Lambda$CDM one and let us focus on objects that collapse at a given redshift, say $z=0$. The initial overdensity must be rather bigger in the $\Lambda$CDM case to overcome the acceleration induced by $\Lambda$. But the threshold $\delta_c$ is obtained evolving this initial value with the linear transfer functions and this will suppress the $\Lambda$CDM value, exactly for the same reason which required it to be bigger in the first place. In other words the only effect comes from the difference between the linear and non-linear evolution and this causes only a small suppression with respect to the CDM case.  

\section{The mass function of dark matter halos}
\label{sec:massfunction}

We are now ready to discuss the predictions for the mass function using the Press-Schechter formalism \cite{Press:1973iz,Bond:1990iw}.  We will first concentrate on the mass of dark matter, leaving aside for the moment the contribution of the quintessence mass to the halos. The volume density of dark matter halos of mass $M$ is given by
\be
\label{PSmassfunction}
\frac{dn_{\rm PS}}{dM}(M,z) = - \sqrt\frac{2}{\pi} \frac{\bar \rho_m}{M^2} \frac{\delta_c(z)}{D(z) \sigma_M} \frac{d \log \sigma_M}{d \log M} \exp\left[-\frac{\delta^2_c(z)}{2 D^2(z)\sigma_M^2}\right] \;.
\ee
Here $\sigma_M^2$ is the smoothed variance of the density field which we define with a sharp cut-off in real space
\be
\sigma_M^2 \equiv \frac1{2 \pi^2} \int_0^{\infty} d k \; k^2 |W(kR) |^2 P_m(k)  \qquad {\rm with} \qquad M \equiv \frac{4 \pi}{3} R^3 \bar\rho_m \;,
\ee 
where $P_m$ is the matter power spectrum and $W(kR) \equiv 3 (\sin kR - kR \cos k R)/ (kR)^3$ is the Fourier representation of the top-hat window function in real space. 
Note that the Press-Schechter mass function (\ref{PSmassfunction}) can be rigorously derived only using a sharp filter in Fourier space. Thus its use with a sharp filter in real space is just an approximation; for corrections to this approximation see for example \cite{Maggiore:2009rv}. Notice that the redshift dependence of the threshold $\delta_c(z)$ only comes from the spherical collapse dynamics discussed in the previous Section and does not include the growth of the matter power spectrum, which is separately taken into account by the linear growth function $D(z)$. 

\begin{figure}[!!!h]
\begin{center}
\includegraphics[scale=1.2]{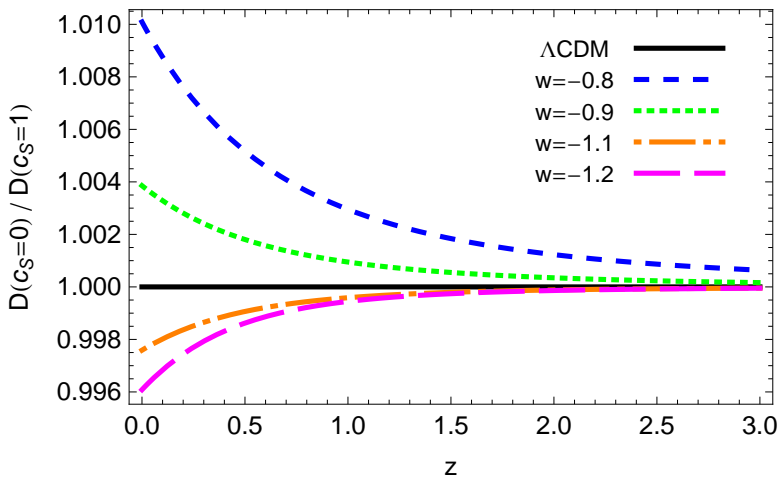}  
\end{center}
\caption{\small {\em Ratio between the growth functions $D(z)$ for $c_s=0$ and $c_s=1$ as a function of the redshift $z$.}}
\label{fig:growth}
\end{figure}

The linear matter power spectrum is very similar in the cases $c_s =1$ and $c_s =0$. The difference comes from the contribution to the Poisson equation of quintessence perturbations as shown in eq.~\eqref{master4}. The effect is independent of $k$ and it is thus equivalent to a change in the growth function $D(z)$. The change in the growth function can be easily calculated as a function of $z$ by solving eqs \eqref{master4} and \eqref{master5}. The result is shown in fig.~\ref{fig:growth}. Given that quintessence becomes relevant only recently and that perturbations are suppressed by $1+w$ the effect on dark matter does not exceed the percent level; this result is somewhat smaller than a naive estimate one can get by comparing the two contributions on the right-hand side of eq.~\eqref{master4}. As one can see in fig.~\ref{fig:growth}, for $1+w > 0$ setting the speed of sound of quintessence to zero fosters the clustering and the dark matter spectrum is slightly enhanced; for $1+w <0$ quintessence clustering has negative energy and the dark matter spectrum is suppressed. The size of the effect is smaller for  $1+w <0$ as quintessence becomes relevant only very recently.

\begin{figure}[!!!h]
\begin{center}
\includegraphics[scale=1.4]{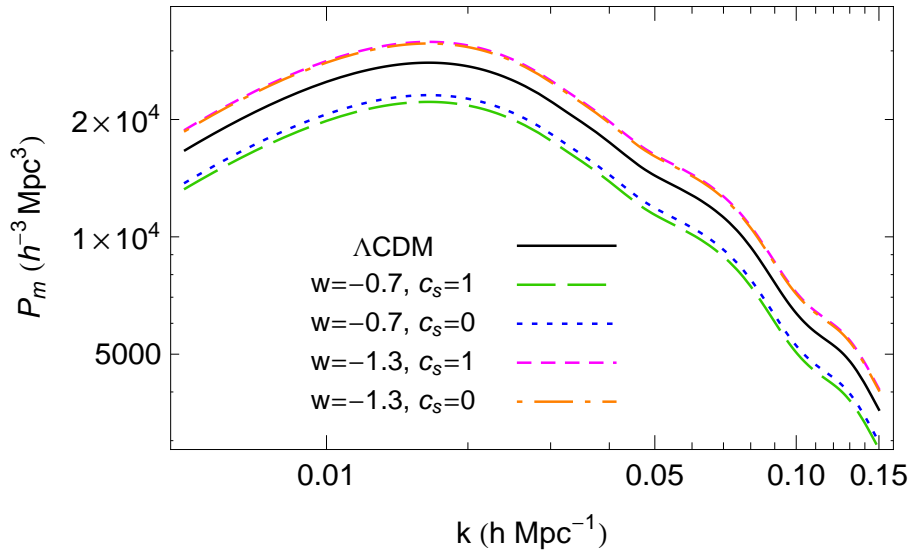}  
\end{center}
\caption{\label{fig:matterpower} \small {\em Matter power spectrum. The two lines with $w=-1.3$ are almost superimposed.}}
\end{figure}

For the calculation of the mass function, we do not need only the growth rate, but the complete matter power spectrum. We use for this the publicly available code CAMB \cite{Lewis:1999bs}, which allows to set to zero the speed of sound of quintessence. Apart from $w$ all the other cosmological parameters are set to the WMAP5 best fit values \cite{Komatsu:2008hk}. We have checked that the effect of setting $c_s=0$ instead of $c_s=1$ in the code is compatible with what we got in fig.~\ref{fig:growth}. In figure \ref{fig:matterpower} we show the matter power spectrum with $c_s=0$ and $c_s=1$ for two different values of $w$. The speed of sound gives a small effect, much smaller than the modification of the growth rate induced by the different background: for $w > -1$ quintessence becomes relevant before and suppresses the spectrum in comparison with $\Lambda$CDM. The opposite effect is obtained in the case $w < -1$.

The equation of state $w$ and the speed of sound of quintessence enter in the mass function eq.~\eqref{PSmassfunction} modifying the growth function $D(z)$ and the threshold for collapse $\delta_c(z)$. Notice that only the combination $\delta_c(z)/D(z)$ enters in the Press-Schechter formula. As we discussed in the previous Section, the change in the threshold is very suppressed and much smaller than the correction of the growth function. 

\begin{figure}[!!!h]
\begin{center}
\includegraphics[scale=1.2]{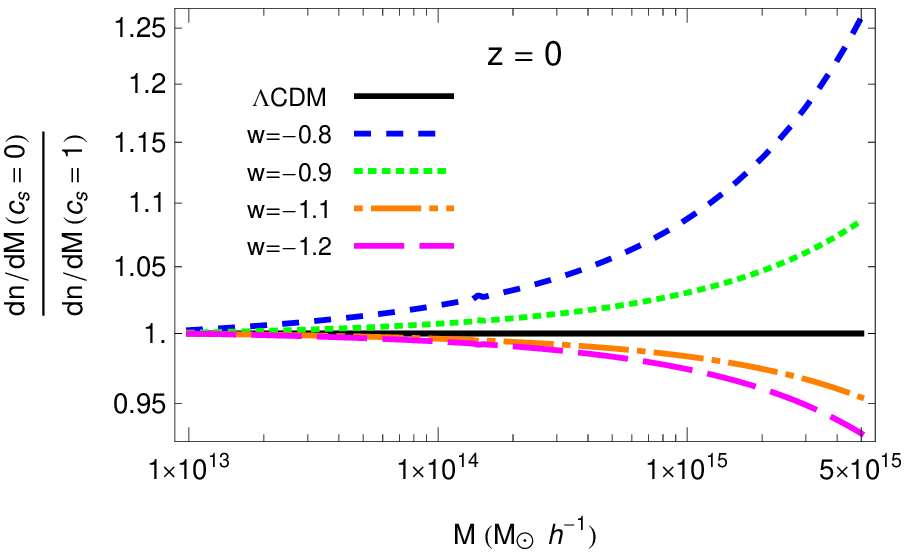}  

\vspace{0.5cm}
\includegraphics[scale=1.2]{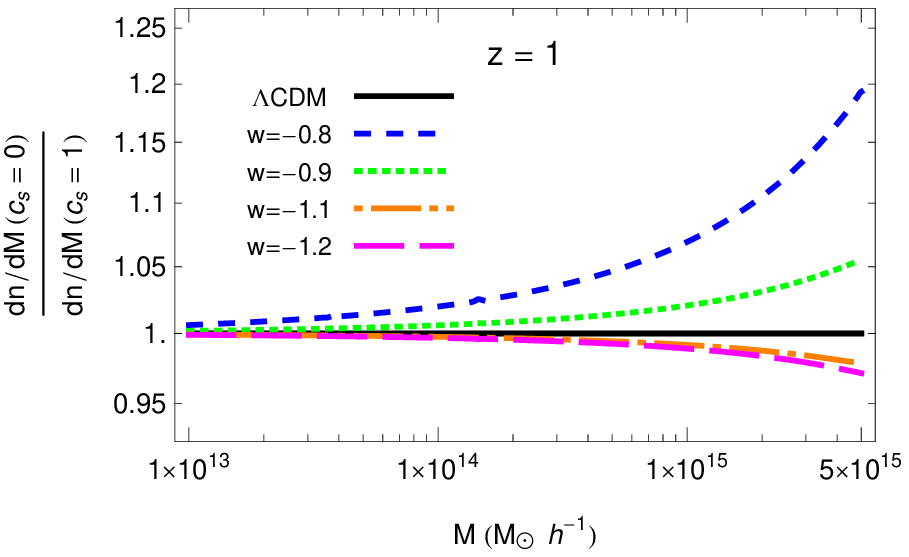} 
\end{center}
\caption{\label{fig:massfunctionratio}\small {\em Ratio of the Press-Schechter mass function for $c_s=0$ and $c_s=1$ at $z=0$ (above) and $z=1$ (below).}}
\end{figure}

We want to focus on the effect of clustering quintessence on the collapse with respect to the case when quintessence remains unperturbed. This effect can be estimated taking the ratio of the Press-Schechter mass functions in the two cases,
\be
\frac{d n_{\rm PS}/dM (w, c_s=0)}{d n_{\rm PS}/dM (w, c_s=1)} \;,
\ee
plotted in figure \ref{fig:massfunctionratio}. We see that the effect is quite small: the ratio becomes large at high mass as a consequence of the exponential dependence of the number density on the mass. We do not dwell on the measurability of this small effect because, as we will discuss in the next section,  the additional contribution to the mass of the halo coming from the clustered quintessence will give a comparable change in the mass function.

\begin{figure}[!!!h]
\begin{center}
\includegraphics[scale=1.1]{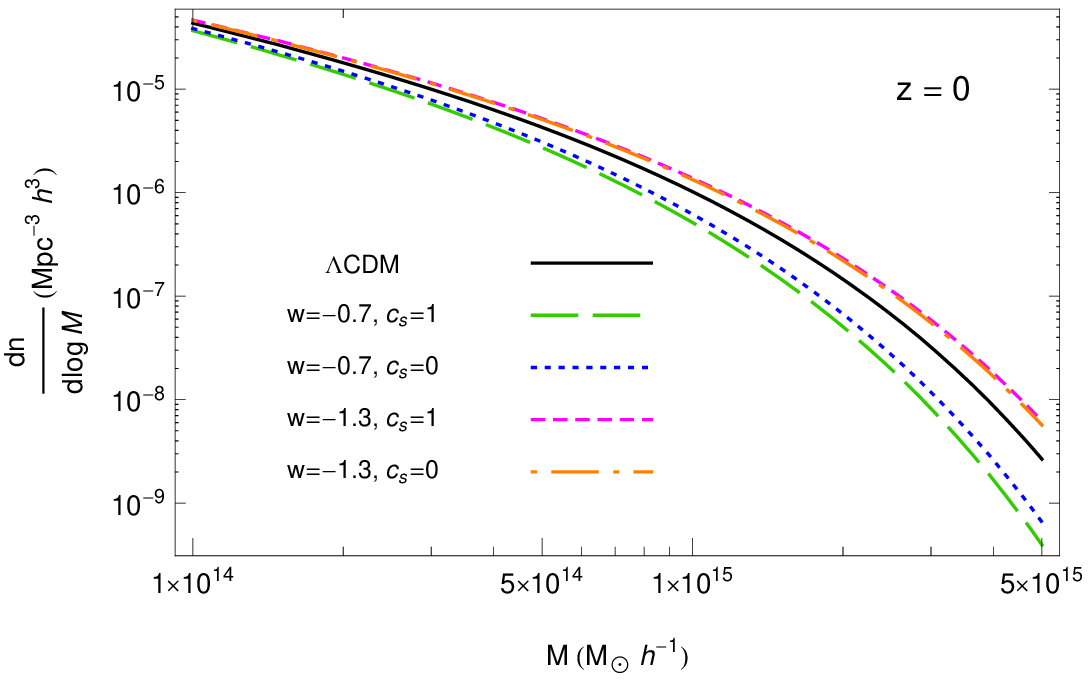}  

\vspace{0.5cm}
\includegraphics[scale=1.1]{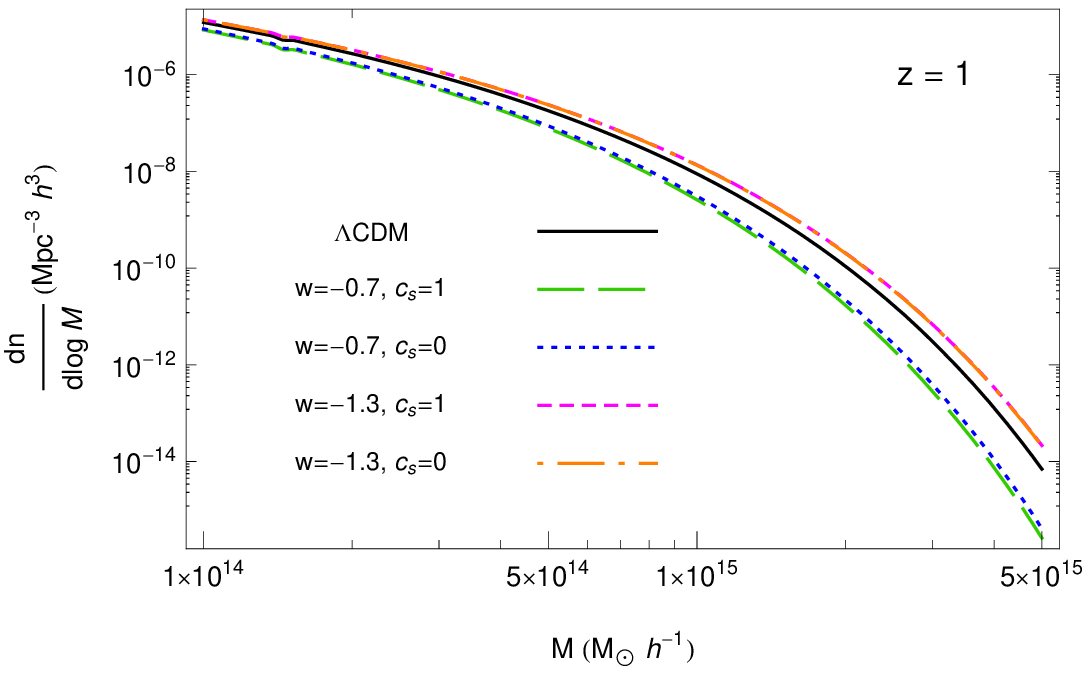}
\end{center}
\caption{\label{fig:massfunction}\small {\em Mass function calculated using eq.~\eqref{STratio} for $z=0$ (above) and $z=1$ (below).}}
\end{figure}

It is well known that the Press-Schechter formula does not fit in detail the mass function obtained by numerical simulations. A better fit, motivated by the ellipsoidal collapse model, is given by the Sheth-Tormen mass function \cite{Sheth:1999mn},
\be
\frac{d n_{\rm ST}}{d M}(M,z) = -\sqrt\frac{2 a}{\pi} A \left[1+\left(\frac{a \delta_c^2}{D(z)^2 \sigma_M^2}\right)^{-p}\right] \frac{\bar \rho}{M^2} \frac{\delta_c}{D(z) \sigma_M} \frac{d \log \sigma_M}{d \log M} \exp\left[-a \frac{\delta_c^2}{2 D(z)^2 \sigma_M^2}\right]\;,
\ee
with $a=0.707$, $A =0.322184$ and $p=0.3$. Since, as discussed, the dependence of the threshold $\delta_c$ on the cosmology  is very mild in all cases, $\delta_c$ is usually taken to be $z$ independent and equal to the EdS value, $\delta_c =1.686$. Notice that in this way the function is ``universal'' in the sense that the dependence on the cosmological parameters and redshift is only through the smoothed linear density field $D(z) \sigma_M$.   It is reasonable to expect that the Sheth-Tormen formula gives a good description of the mass function in the case of non clustering quintessence. In this case the only effect of quintessence is through the time dependence of the background and its effect on the growth function; this is not qualitatively different from the case of $\Lambda$CDM. In other words we expect ``universality'' to hold also on this case. On the other hand, we can estimate the effect of clustering quintessence using the ratio described above: in taking the ratio we expect that the shortcomings of the Press-Schechter prescription will partially cancel. Therefore, in figure \ref{fig:massfunction} we plot
\be
\label{STratio}
\frac{d n}{d M} \equiv \frac{d n_{\rm ST}}{d M}(w,c_s=1)  \cdot \frac{d n_{\rm PS} /dM (w, c_s=0)}{d n_{\rm PS} /dM (w, c_s=1)} \;.
\ee
As expected the main effect is at low redshift and high masses.

\section{Quintessence contribution to the halo mass}
\label{sec:quintmass}

So far we have been interested in the contribution of dark matter to the halo mass function. Given that quintessence with vanishing speed of sound participates in the collapse, one  may wonder whether   
quintessence will contribute a sizable amount of mass to the dark matter halo. After all, most of the measurements will be sensitive to the total mass of the object, and not only to the fraction of it associated with dark matter. 

\begin{figure}[!!!h]
\begin{center}
\includegraphics[scale=1]{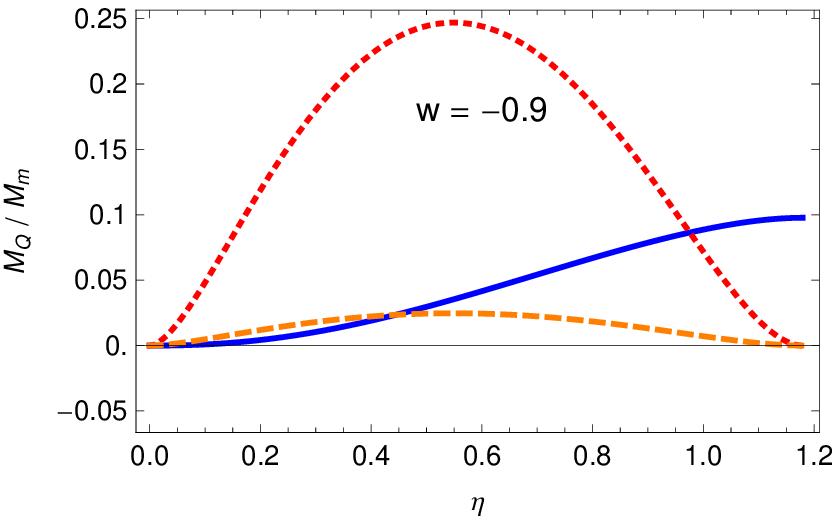}  
\vspace{0.5cm}

\includegraphics[scale=1]{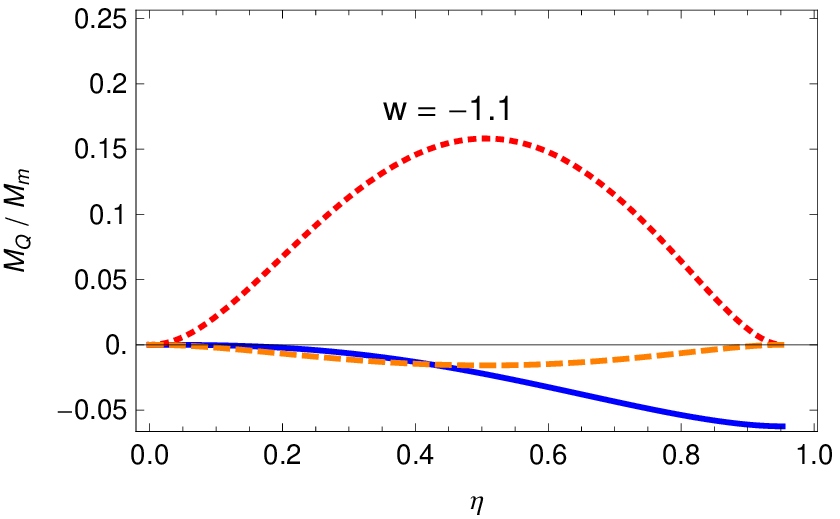}
\end{center}
\caption{\label{fig:ratiosph}\small {\em Spherical collapse for $w=-0.9$ (above) and $w=-1.1$ (below). Solid blue line: the quintessence contribution to the halo mass $(4\pi/3)R^3 \delta\rho_Q$, normalized to the constant dark matter mass, as a function of time. Note that it becomes constant at late time. Red dotted line: the same, but for the background energy density $\bar\rho_Q$. Orange dashed: the same for $(1+w) \bar\rho_Q$.}}
\end{figure}

The quintessence contribution to the halo mass can be defined as
\be
\label{extramass}
M_Q \equiv \int_{\rm M} \!\ud^3 x \;\delta\rho_Q \;,
\ee
where the integral is extended over the whole dark matter overdensity. If we stick to the spherical collapse model this just reduces to $(4\pi / 3)R^3 \delta\rho_Q$.  Of course it makes sense to interpret this expression as a contribution to the halo mass, only if it stays practically constant over the time scales of interest, i.e.~a Hubble time.
In the spherically symmetric case $\delta\rho_Q$ follows equation \eqref{deltarhoEOM},
\be
\dot{\delta \rho}_Q + 3\frac{\dot R}R \delta \rho_Q = - {3 \delta H}{}(\bar \rho_Q + \bar p_Q)\,.
\ee
When $|\delta\rho_Q| \gg |1+w| \bar\rho_Q$, the right-hand side is negligible. In this limit $\delta\rho_Q$ redshifts as matter so that the integral \eqref{extramass} becomes constant as can be seen in figure \ref{fig:ratiosph}. We expect this condition to be marginally satisfied at turn-around when $\delta\rho_Q \sim (1+w) \bar\rho_Q$.  This allows us to estimate the quintessence contribution to the halo mass,
\be
\frac{M_Q}{M_{m}} \sim (1+w) \frac{\Omega_Q}{\Omega_m} \;,
\ee
although for a precise estimate one cannot neglect the evolution of the quintessence mass after turn-around.
Notice also that quintessence with $1+w <0$ contributes with a negative mass.

Of course this is only the prediction of the idealized spherical collapse solution. In reality dark matter halos virialize with an overdensity $\sim200$ times larger than the background. What happens to quintessence while dark matter virializes? Quintessence is exactly comoving with dark matter;  eventually dark matter reaches shell crossing and the velocity field ceases to be single-valued. This corresponds to the formation of cusps in the quintessence field, similarly to what discussed in the ghost condensate case \cite{ArkaniHamed:2005gu} and more recently in the context of Ho\v{r}ava-Lifshift gravity \cite{Mukohyama:2009mz, Blas:2009yd}.  The dynamics of the cusps will depend on higher derivative operators and possibly on the UV completion of the theory.\footnote{For positive $1+w$ one has $c_s^2>0$. In this case, one would naively expect that quintessence remains smooth on very short scales thus preventing the formation of caustics. However, in our case the velocity of the quintessence fluid (which is the same as the dark matter velocity) exceeds the speed of sound, i.e.~it is ``supersonic''. In this case sound waves are too slow to prevent the formation of caustics.} In any case, the dynamics of quintessence in this phase is very complicated \cite{ArkaniHamed:2005gu} and its treatment is beyond the scope of this paper. Let us assume however that the cusps are somehow regularized and try to draw some general conclusion that is independent of the details of virialization.\footnote{We are implicitly assuming that the process responsible for smoothing the cusps does not lead to an energy loss to infinity. For instance, this happens if the smoothing excites new relativistic degrees of freedom which are radiated away.} 

\begin{figure}[!!!h]
\begin{center}
\includegraphics[scale=1.2]{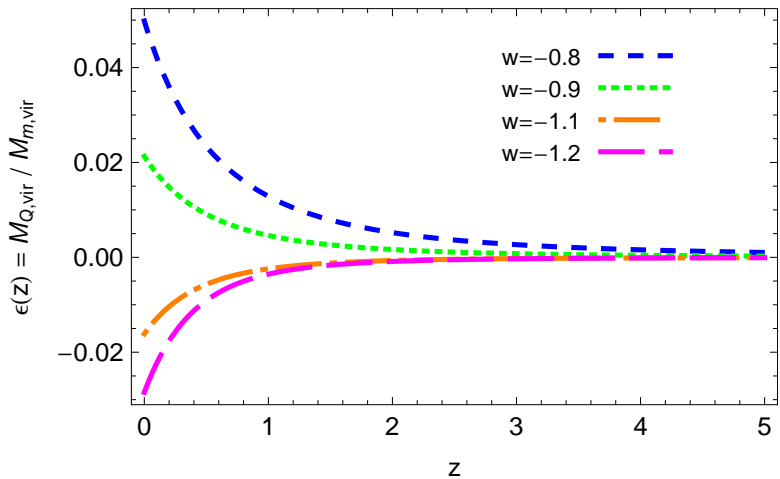}  
\end{center}
\caption{\label{fig:ratiovir}\small {\em The mass contribution of quintessence to the total halo mass, calculated from the spherical collapse solution when the radius reaches the virialization value. }}
\end{figure}

As $\delta\rho_Q \simeq (1+w) \bar\rho_Q$ at turn-around, when matter starts virialization the inequality $|\delta\rho_Q| \gg |1+w| \bar\rho_Q$ is satisfied with rather good approximation: as the virial radius is approximately half of the turn-around radius, $\delta\rho_Q$ has grown by approximately 8 times from the beginning of the collapse while $\bar\rho_Q$ has remained approximately constant. The spherical collapse solution indicates that the quintessence mass is with good approximation constant at virialization. In the real case we expect the variation of the quintessence mass to be even smaller;  the conservation of the stress-energy tensor tells us that the mass variation is related to the energy flux across a surface around the object,
\be
\dot M = \oint d S_i \;T_0^i \;.
\ee
As during virialization the velocity field of quintessence will cease to be radial, we expect this integral to be suppressed with respect to the spherically symmetric case.
We conclude that, as the flux integral is negligible, {\em the mass associated to quintessence stays constant independently of the details of virialization}.  
Thus, a good estimate of the quintessence mass can be obtained from the spherical collapse model evaluating the quintessence contribution at the virialization radius, see figure \ref{fig:ratiovir}. The ratio between the virialization and the turn-around radii is taken to be the same as in $\Lambda$CDM \cite{Lahav:1991wc}.\footnote{Quintessence clustering will modify the virial radius with respect to $\Lambda$CDM with corrections ${\cal{O}}(|1+w|)$. The effect of these corrections on the amount of clustered quintessence is ${\cal{O}}(|1+w|^2)$ and can therefore be safely neglected.}

Let us see how we can take into account the additional quintessence mass in the prediction for the mass function. A rigorous treatment would be quite challenging: the formation of an object of mass $M$ at redshift $z$ should be accompanied by an extra mass $M \to M [ 1+ \epsilon(z) ]$, where $\epsilon(z)\equiv M_{Q,{\mathrm vir}} / M_{m, {\mathrm vir}}$ is the quintessence to dark matter mass ratio at virialization. Then one should follow this extra mass as the halo merges to form larger objects which in turn accrete extra quintessence as they form.  However $\epsilon(z)$ is important only at low redshifts,  so that we expect the main effect to be on large objects that formed very recently, the ones on the exponential tail of the mass function. Let us see how we can estimate the effect in this region. At any epoch, the largest objects are mostly created with a negligible rate of destruction through merging to form larger objects. Therefore the formation rate for these objects can be accurately approximated by
\be
- \frac{\partial}{\partial z} \frac{d n_{\rm PS}}{d \log M}(M,z) \;.
\ee
As the effect of quintessence is to rescale the mass of the object as it forms, it is more convenient to use the mass function per logarithmic mass interval.  
Using this expression, when positive, as an approximation for the formation rate gives 
\be
\label{PSQ2}
\frac{\partial }{\partial z} \frac{dn_{{\rm PS}, m + Q}}{d \log M}(M,z) = \frac{\partial }{\partial z} \frac{dn_{\rm PS}}{d \log M}\left(M(1-\epsilon(z)),z\right) \;.
\ee
Expanding the right-hand side of this expression and integrating it over the redshift
enables us to take into account the extra mass associated with quintessence accreted from an inizial redshift $z_i$,
\be
\begin{split}
\label{PSQ}
\frac{dn_{{\rm PS}, m + Q}}{d \log M}(M,z) & =   \frac{dn_{\rm PS}}{d \log M}(M,z) \\ & +  \int^{z}_{z_{i}} d\tilde z \;\epsilon(\tilde z) \left[- \frac{\partial}{\partial \log M} \frac{\partial}{\partial \tilde z} \frac{dn_{\rm PS}}{d \log M}(M, \tilde z)\right]  \cdot \theta \left( -  \frac{\partial}{\partial \tilde z} \frac{dn_{\rm PS}}{d \log M}(M, \tilde z) \right) \;,
\end{split}
\ee
where $\theta(x)$ is the Heaviside theta function.

\begin{figure}[!!!h]
\begin{center}
\includegraphics[scale=1.1]{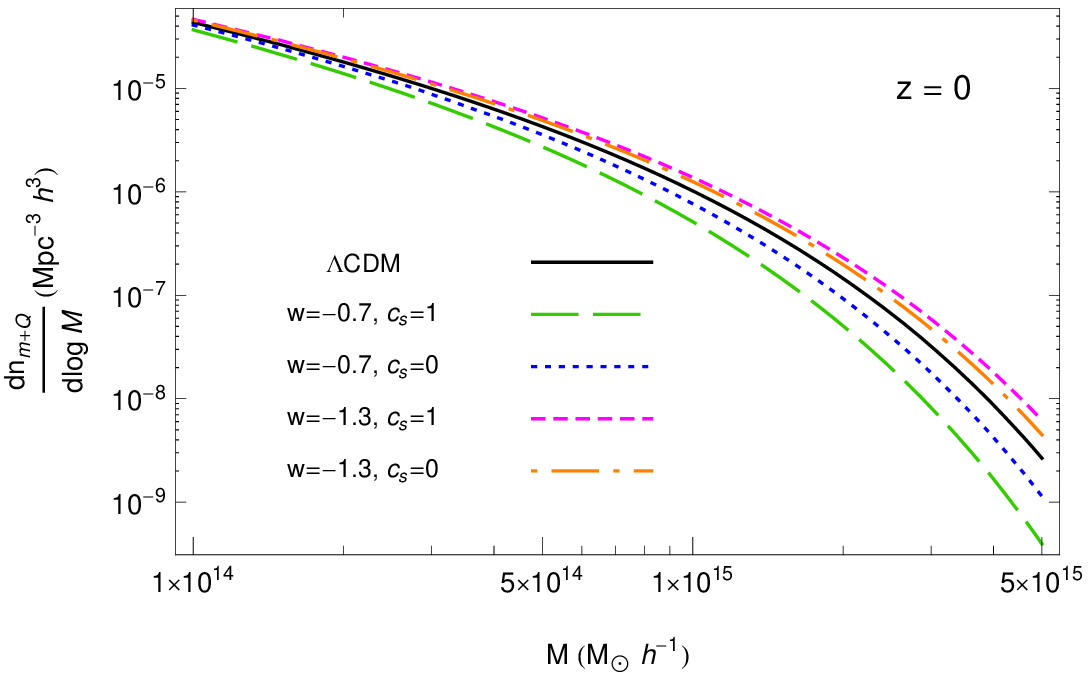}  

\vspace{0.5cm}
\includegraphics[scale=1.1]{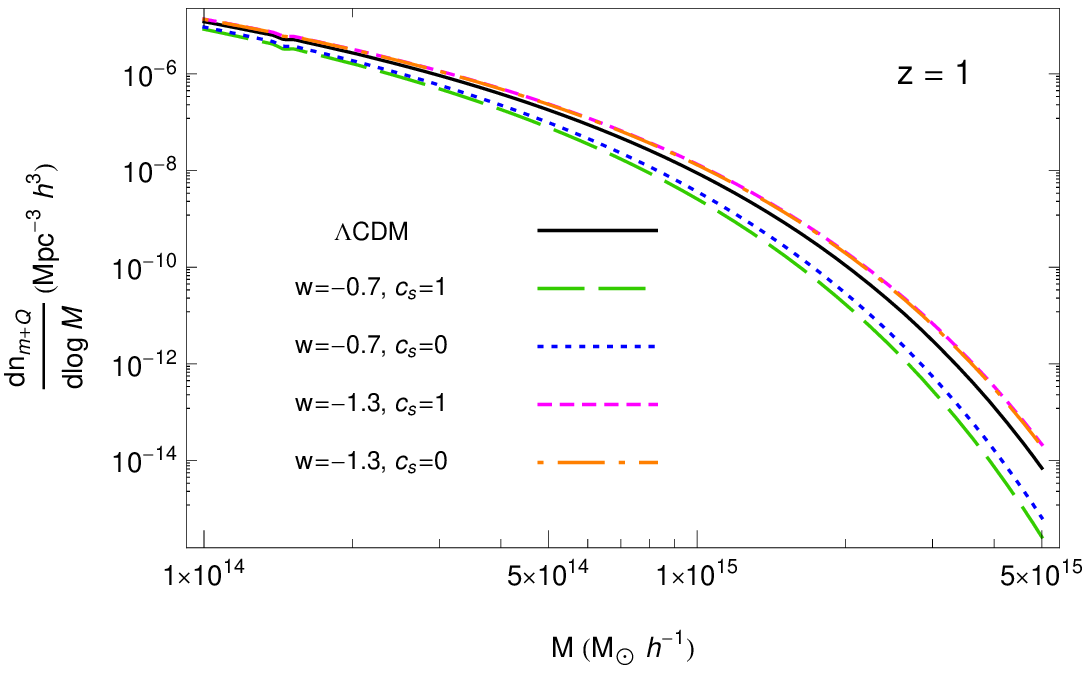}
\end{center}
\caption{\label{fig:qmassfunction}\small {\em Mass function for $z=0$ (above) and $z=1$ (below), including the quintessence mass contribution, calculated using eq.~\eqref{PSQ} and \eqref{STratio}.}}
\end{figure}

\begin{figure}[!!!h]
\begin{center}
\includegraphics[scale=1.2]{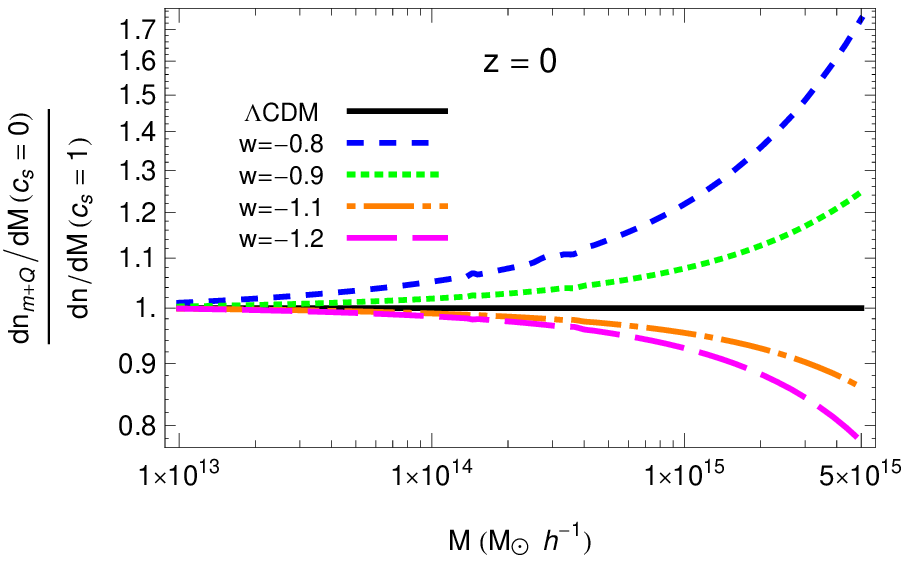}  

\vspace{0.5cm}
\includegraphics[scale=1.2]{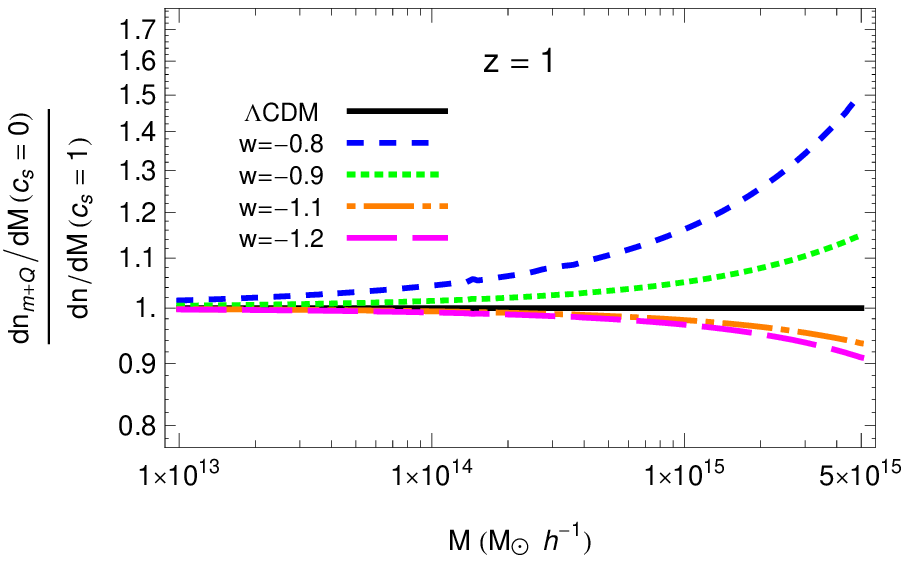} 
\end{center}
\caption{\label{fig:qmassfunctionratio}\small {\em Ratio of the Press-Schechter mass function, including the quintessence contribution eq.~\eqref{PSQ}, for $c_s=0$ and $c_s=1$ at $z=0$ (above) and $z=1$ (below).}}
\end{figure}

In figure \ref{fig:qmassfunction} we plot the mass functions including the quintessence mass contribution, using the new Press-Schechter mass formula eq.~\eqref{PSQ} into eq.~\eqref{STratio}. 
Notice that the effect of quintessence mass is to bring the $c_s=0$ lines closer to the $\Lambda$CDM one.
To better visualize this effect, in figure \ref{fig:qmassfunctionratio} we plot the ratio between the $c_s=0$ and $c_s=1$ case. For $w >-1$ ($w < -1$), setting to zero the speed of sound of quintessence not only does it enhance (diminish) the clustering of dark matter as discussed in the previous Sections, but it also adds positive (negative) mass to the halo: the two effects therefore pile up and the second is quantitatively slightly dominant.  The sum of the two effects is rather large: for values of $w$ still compatible with the present data and for large masses the difference between the predictions of the $c_s = 0$ and the $c_s = 1$ cases is of order one.

We stress that the new mass function is not universal in the sense that there is an explicit red-shift dependence besides the one implicit in the growth of $\sigma_R$  (\footnote{The universality of the mass function, even in the presence of scale independent non-Gaussianity, has been tested with good accuracy in $N$-body simulations \cite{Pillepich:2008ka}.}). This $z$ dependence is quite remarkable: the fact that the modification takes place only at very low red-shift is quite distinctive and a clear link of the effect with the onset of acceleration.

\section{Three contributions to the mass}
\label{sec:3masses}
In the previous section we saw that a distinctive signature of quintessence with $c_s=0$ is the extra contribution of quintessence to the mass of virialized objects.\footnote{In this section we will assume this mass to be positive, which is the case for $1+w >0$. Similar considerations apply to the $1+w <0$ case.} Although a detailed study of how to distinguish this extra contribution goes beyond the scope of this paper, few remarks are in order. Let us focus on clusters: as discussed in the previous section, the mass of these large objects is significantly affected by the quintessence mass. Moreover, clusters
are mostly dominated by gravitational physics (for an introduction to the subject see \cite{Borgani:2006ba}). If we neglect quintessence for the moment, a cluster is characterized by its baryon mass, mostly in the form of gas, and by the dark matter mass. For sufficiently large clusters, in first approximation one expects the ratio between these two components to be close to the cosmological baryon to dark matter ratio \cite{White:1993wm}. 

The various techniques to study clusters have different sensitivities to the two components (for a review on cluster detection tecniques see \cite{Voit:2004ah}). Weak lensing and galaxy velocity measurements are sensitive to the gravitational mass of the object, i.e.~to the sum of the dark matter and baryon mass. On the other hand the optical richness, a measure of the light emitted by galaxies inside the cluster, is a probe of the baryon fraction of the mass. X-ray and Sunyaev-Zeldovich (SZ) measurements are sensitive both to the dark matter component and to the baryonic one. While these effects are caused by the gas in the cluster, the temperature and the equilibrium of the intergalactic medium depends on the dark matter distribution. Indeed, assuming hydrostatic equilibrium one can use the X-ray image of a cluster to reconstruct from its temperature and luminosity distribution the baryonic and dark matter radial profiles, as explained for instance in \cite{Vikhlinin:2005mp,Mantz:2009fx}.  The combination of all these techniques should allow a robust reconstruction of the dark matter and baryon components of the clusters. 

These standard techniques assume that there is no extra contribution to the cluster mass. However, the crucial property of $c_s =0$ models of quintessence is that a third kind of mass, associated to quintessence, enters in the game. How are the different tecniques sensitive to these new mass? Lensing and galaxy velocity measurements will still trace the total mass, now including, after dark matter and baryons, also quintessence.  On the other hand, optical light associated to galaxies should be rather independent of the accreted quintessence.  X-ray experiments should indicate a higher gravitational mass -- both from studying the hydrostatic equilibrium and from temperature measurements which are sensitive to the gravitational potential well -- but the same baryon component. This effect is similar to a reduction of the $f_{\rm gas}$ parameter, the ratio between baryonic and dark matter mass. Similar considerations apply also to SZ measurements.

A useful lever arm for distinguishing the accretion of dark energy from the other uncertainties in the description of a cluster is the strong red-shift dependence. Quintessence should be relevant only at very low redshift. It would be useful to find out the best combination of observables which is able to constrain the presence of extra mass in the cluster. Besides a smoking gun of clustering quintessence, it would be a useful consistency check between the various measurements.

\section{Conclusions}
\label{sec:conclusions}


Using the spherical collapse model, we have studied how the clustering of quintessence with negligible speed of sound can affect the prediction for the mass function of dark halos. As quintessence does not develop pressure gradients, it follows geodesic motion remaining comoving with the dark matter. In contrast to the case where quintessence remains smooth, spherical collapsing regions behave as exact closed FLRW solutions with the quintessence contributing to the overdensity. To study the spherical collapse we found it useful to use Fermi coordinates where the effect of the expansion can be treated as a perturbation around Minkowski spacetime locally around a spatial point. In contrast to comoving coordinates, a unique coordinate system can be employed to describe both the interior and exterior of the collapsing region. 



Quintessence with zero speed of sound modifies dark matter clustering with respect to the smooth quintessence case through the linear growth function and the linear threshold for collapse. Besides these conventional effects there is a more important and qualitatively new phenomenon: quintessence mass adds to the one of dark matter,
contributing to the halo mass by a fraction of order $(1+w) \Omega_Q/\Omega_m$. This effect is quite difficult to model and in this paper we have adopted a simplified treatment which gives an accurate estimate of the high mass tail of the distribution where the effect is more relevant.

As dark energy plays an active role in the formation of structures, the distinction between what we call dark matter and dark energy becomes fuzzy. The distinction between the two components can be probed by the different redshift dependence.
It is quite remarkable that our knowledge of structure formation still allows a completely inhomogeneous dark energy component on short scales.

In this paper we did not attempt to study experimental constraints and forecasts. However, we have seen that for values of $w$ which are not too close to the cosmological constant one, let's say $|1+w| \gtrsim 0.1$, the predictions for the high mass tail of the mass function in the $c_s=0$ and the $c_s=1$ cases are quite distinctive. The effect of clustering quintessence gives order one changes in the expected number of clusters. Whether the effect will be measurable or not will depend on the possibility of breaking the degeneracy with cosmological parameters, most notably  $\sigma_8$ and $\Omega_m$, and on the good recostruction of the limiting mass of the survey. Similar concerns have been addressed in the context of the effects of primordial non-Gaussianity on the mass function in \cite{LoVerde:2007ri}.

Our work can be continued in various directions.  A better theoretical treatment of quintessence accretion would strengthen our predictions for the mass function, although at a certain point only a (challenging) numerical simulation can make a fully reliable prediction. On the data side it would be interesting to understand what is the minimum value of $|1+w|$ that allows a distinction between $c_s=0$ and $c_s=1$ using the forthcoming cluster mass function measurements. One should also explore whether other probes, lensing for example, are better suited to study the non-linear clustering of quintessence. It would also be interesting to explore the effect of quintessence on the halo dynamics, but this probably requires a way to deal with the formation of caustics. In a more model-independent way, one could try to use data on clusters to constrain the presence of extra gravitational mass besides the dark matter mass. We leave all this for future work.

\section*{Acknowledgements}
It is a pleasure to thank Steve Allen, Pierstefano Corasaniti, Eric Linder, Marcello Musso, Aseem Paranjape, David Rapetti, Graziano Rossi, Sarah Shandera, Ravi Sheth, Anze Slosar, Patrick Valageas, Risa Wechsler and Matias Zaldarriaga for useful discussions. J.N.~and F.V.~ackowledge support from the EU Marie Curie Research and Training network ``UniverseNet'' (MRTN-CT-2006-035863).

\end{document}